\documentclass[12pt, draftclsnofoot, onecolumn]{IEEEtran}
\usepackage{amsmath,amssymb,amsfonts,amsthm}
\usepackage[square, comma, sort&compress, numbers]{natbib}
\usepackage{algorithm, algpseudocode}
\usepackage{array}
\usepackage{multicol}
\usepackage{multirow,booktabs}
\usepackage{graphicx}
\usepackage{graphicx,pgf,tikz}
\usetikzlibrary{arrows,automata}

\ifCLASSOPTIONcompsoc
\usepackage[caption=false,font=normalsize,labelfont=sf,textfont=sf]{subfig}
\else
\usepackage[caption=false,font=footnotesize]{subfig}
\fi

\IEEEoverridecommandlockouts
\theoremstyle{definition}
\newtheorem{theorem}{Theorem}
\newtheorem{lemma}{Lemma}
\newtheorem{corollary}{Corollary}

\allowdisplaybreaks

\begin{document}
\title{Delay-Optimal Buffer-Aware Scheduling with Adaptive Transmission}
\author{
Xiang~Chen,~\IEEEmembership{Student~Member,~IEEE,}
Wei~Chen,~\IEEEmembership{Senior~Member,~IEEE,}
Joohyun~Lee,~\IEEEmembership{Member,~IEEE,}
and~Ness~B.~Shroff,~\IEEEmembership{Fellow,~IEEE}
\thanks{X. Chen and W. Chen are with the Department of Electronic Engineering and Tsinghua National Laboratory for Information Science and Technology (TNList), Tsinghua University. E-mail: chen-xiang12@mails.tsinghua.edu.cn, wchen@tsinghua.edu.cn.

J. Lee is with the Department of ECE at The Ohio State
University. E-mail: lee.7119@osu.edu.

Ness B. Shroff holds a joint appointment in both the Department of ECE and the Department of CSE at The Ohio State University. E-mail: shroff@ece.osu.edu.
}}
\maketitle

\begin{abstract}
In this work, we aim to obtain the optimal tradeoff between the average delay and the average power consumption in a communication system. In our system, the arrivals occur at each timeslot according to a Bernoulli arrival process and are buffered at the transmitter. The transmitter determines the scheduling policy of how many packets to transmit under an average power constraint. The power is assumed to be an increasing and convex function of the number of packets transmitted in each timeslot to capture the realism in communication systems. We also consider a finite buffer and allow the scheduling decision to depend on the buffer occupancy. This problem is modelled as a Constrained Markov Decision Process (CMDP). We first prove that the optimal policy of the (Lagrangian) relaxation of the CMDP is deterministic and threshold-based. We then show that the optimal delay-power tradeoff curve is convex and piecewise linear, where each of the vertices are obtained by the optimal solution to the relaxed problem. This allows us to show the optimal policies of the CMDP are threshold-based, and hence can be implemented by a proposed efficient algorithm. The theoretical results and the algorithm are validated by Linear Programming and simulations.
\end{abstract}
\begin{IEEEkeywords}
Cross-layer design, Joint channel and buffer aware scheduling, Markov Decision Process, Queueing, Energy efficiency, Average delay, Delay-power tradeoff, Linear programming.
\end{IEEEkeywords}

\section{Introduction}
Scheduling for minimizing delay or power has been studied widely and is getting increasingly important, as many delay sensitive applications are emerging, such as instant messenger (IM), social network service (SNS), streaming media and so on.  On the other hand, the requirements of mobility and portability for communication terminals incur stringent energy constraints.

\begin{figure}[t]
\centering
\includegraphics[width=0.5\columnwidth]{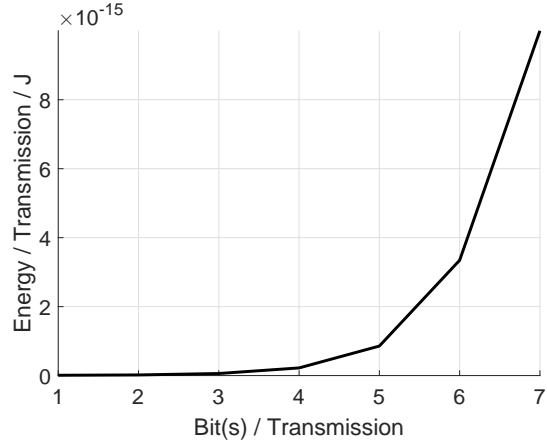}
\caption{Energy~/~Transmission versus Bit~/~Transmission with adaptive M-PSK (Target ber=$10^{-5}$, Noise Power Spectral Density $N_0$=-150 dBm/Hz)}
\label{fig_psk}
\vspace{-0.8cm}
\end{figure}

In typical communication systems, for fixed channel conditions, the power efficiency (per bit transmitted) rapidly decreases as the transmission rate is increased. In other words, the power cost is convex in transmission rate. Below are two canonical examples of communication systems that demonstrate this convex behaviour.
\begin{enumerate}
\item The information-theoretically optimal transmission rate $R=\frac{1}{2}\log_2(1+\frac{P}{N})$. Therefore the power to transmit $s$ bit(s) is $P_s=N(4^s-1)$, which is strictly increasing and convex.
\item Consider an adaptive M-PSK transmission system with a fixed bit error rate (ber). The ber expression for M-PSK is shown in \cite[(8.31)]{simon2005digital}. We fix the ber=$10^{-5}$ and the one-sided noise power spectral density $N_0$=-150 dBm/Hz. The energy-bit curve is shown in \figurename~\ref{fig_psk}, which is strictly increasing and convex.
\end{enumerate}

The convexity of power cost in transmission rate brings a natural trade-off between power and delay. As we increase the transmission rate, the delay becomes shorter with the cost of low power efficiency, and vice versa. Our main goal is to characterize the optimal delay-power trade-off and obtain an optimal scheduling policy for a given average power constraint.

The optimal delay-power tradeoff and the optimal scheduling policy in the point-to-point communication scenario have been studied in \cite{collins1999transmission, berry2002communication, goyal2003power, bettesh2006optimal, berry2013optimal, rajan2004delay, agarwal2008structural, djonin2007mimo, ngo2009optimality, ngo2010monotonicity, cheung2015dawn}, under the convexity assumption for power cost. Among these works, the power cost is modeled based on Shannon's formula in \cite{berry2002communication, goyal2003power, bettesh2006optimal, berry2013optimal, rajan2004delay}. Since there is no interference in the point-to-point scenario, the power cost is convex in the transmission rate (bits/transmission), similar to the information-theoretical example we introduced above. Lagrange multiplier method has been applied in these works, in order to transform the constrained optimization to unconstrained optimization to simplify the problem. Based on this, the properties of the delay-power tradeoff curve have been studied in \cite{berry2002communication, berry2013optimal, rajan2004delay, djonin2007mimo}, and the monotonicity of the optimal scheduling policy is investigated in \cite{collins1999transmission, goyal2003power, berry2013optimal, agarwal2008structural, djonin2007mimo, ngo2009optimality, ngo2010monotonicity, cheung2015dawn}. However, most papers neither go back to the original constrained problem, nor prove the equivalence between the original and the Lagrangian relaxation problems. Only in \cite{goyal2003power, djonin2007mimo, ngo2009optimality, ngo2010monotonicity}, properties of the optimal policy for the constrained problem are tackled based on the results from the unconstrained problem. However, in \cite{goyal2003power}, the power cost is fixed by Shannon's formula, thus the results cannot be applied to more generalized power models. In \cite{djonin2007mimo}, necessary properties for proof such as the unichain property of policies, the multimodularity of costs, and stochastically increasing buffer transition probabilities are not proved, but just assumed to be correct. In \cite{ngo2009optimality, ngo2010monotonicity}, binary control is considered, i.e., the scheduler only determines to transmit or not to transmit.

We studied the optimal scheduling in \cite{chen2007optimal,chen2015joint,chen2015delay}, considering a single-queue single-server system with fixed transmission rate, and obtained analytical solutions. Interestingly, in these cases, the monotonicity of the optimal policy can be directly obtained by steady-state analysis of the Markov Process and linear programming formulation. Similar approaches have been applied in \cite{jing2010delay-minimal,saleh2013cross}. We generalized our model and included the adaptive transmission assumption in \cite{chen2015adaptive}, which is much harder to analyse because of the more complicated state transition of the Markov chain. In this paper, we continue this line of research, analyse the problem within the CMDP framework, and present our thorough analysis and results. We first consider its Lagrangian relaxed version. In the unconstrained MDP problem, we prove that the optimal policy is deterministic and threshold-based. Then, in the CMDP problem, we fully characterize the optimal power-delay tradeoff. We prove that the tradeoff curve is convex and piecewise linear, whose vertices are obtained by the optimal policies in the relaxed problem. Moreover, the neighbouring vertices of the trade-off curve are obtained by policies which take different actions in only one state. These discoveries enable us to show that the solution to the overall CMDP problem is also of a threshold form, and devise an algorithm to efficiently obtain the optimal tradeoff curve.

The remainder of this paper is organized as follows. The system model is described in Section II, where the delay-power tradeoff problem is formulated as a Constrained Markov Decision Process. In Section III, based on the Lagrangian relaxation of the CMDP problem, it is proven that the optimal policy for the average combined cost is deterministic and threshold-based. Steady-state analysis is conducted in Section IV, based on which we can prove the optimal delay-power tradeoff curve is piecewise linear, and the optimal policies for the CMDP problem are also threshold-based. Moreover, we propose an efficient algorithm to obtain the optimal delay-power tradeoff curve, and an equivalent Linear Programming problem is formulated to confirm the theoretical results and the algorithm. Simulation results are given in Section V, and Section VI concludes the paper.

\section{System Model}
\begin{figure}[t]
\centering
\includegraphics[width=0.7\columnwidth]{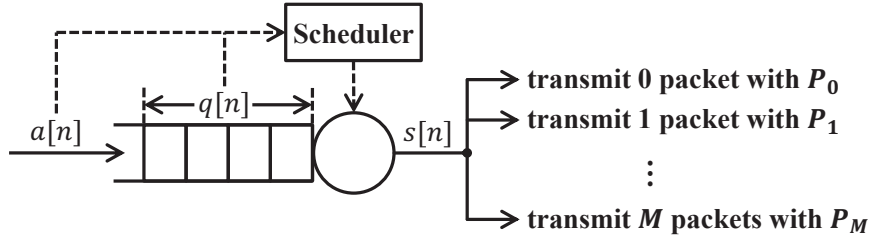}
\caption{System Model}
\label{fig_model}
\vspace{-0.8cm}
\end{figure}

We consider the system model shown in \figurename~\ref{fig_model}. Time is divided into timeslots. Assume that at the end of each timeslot, data packets arrive as a Bernoulli Process with parameter $\alpha$. Each incoming data packet contains $A$ packets. Define $\mathcal{A}=\{0,1\}$. Define $a[n]\in\mathcal{A}$ where $a[n]=1$ or $0$ denote whether or not there are data arriving in timeslot $n$, hence $\text{Pr}\{a[n]=1\}=\alpha$ and $\text{Pr}\{a[n]=0\}=1-\alpha$.

Let $s[n]$ denote the number of data packets transmitted in timeslot $n$. We assume that at most $S$ packets can be transmitted in each timeslot because of the constraints of the transmitter. We force $S \ge A$. Define $\mathcal{S}=\{0,1,\cdots,S\}$, thus $s[n]\in\mathcal{S}$.

Let $p[n]$ denote the power consumption in timeslot $n$. Transmitting $s$ packet(s) incurs power consumption $P_s$, where $0 \le s \le S$. Therefore $p[n]=P_{s[n]}$. Transmitting $0$ packet will cost no power, hence $P_0=0$. Based on analyses in the Introduction, being able to capture the convex relationship between power and bits transmitted is important. Therefore we assume that $P_s$ is strictly increasing and convex in $s$.

The arrivals can be stored in a finite buffer up to a maximum of $Q$ packets. Define $\mathcal{Q}=\{0,1,\cdots,Q\}$. Let $q[n]\in\mathcal{Q}$ denote the number of packets in the buffer at the beginning of timeslot $n$. The amount of transmission $s[n]$ will be decided according to our scheduling policy, based on the historical information of the buffer states and the data arrivals. The dynamics of the buffer is given as
\begin{align}
q[n+1]=q[n]-s[n]+Aa[n].
\label{eqn_q_renew}
\end{align}
To avoid overflow or underflow, the number of transmitted packets in each timeslot $n$, should satisfy $0 \le q[n]-s[n] \le Q-A$.

Consider the queue length $q[n]$ as the state, and the transmission $s[n]$ as the action of the system. According to (\ref{eqn_q_renew}), the transition probability
\begin{align}
\text{Pr}\{q[n+1]=j|q[n]=q,s[n]=s\}=\begin{cases}
\alpha & j=q-s+A,\\
1-\alpha & j=q-s,\\
0 & \text{else}.
\end{cases}
\end{align}
It shows that the probability distribution of the next state is determined by the current state and the chosen action. The queue length $q[n]$ and the transmission power $p[n]$ can be treated as two immediate costs, which are determined by the current state and the current action. Therefore, this system can be considered as a Markov Decision Process (MDP).

A decision rule $\delta_n:\underbrace{\mathcal{Q}\times\mathcal{A}\times\mathcal{Q}\times\cdots\times\mathcal{A}\times\mathcal{Q}}_{n \text{ state(s) and }(n-1)\text{ action(s)}}\rightarrow \mathcal{P}(\mathcal{S})$ specifies action $s[n]$ at timeslot $n$ according to a probability distribution $p_{\delta_n(\cdot)}(\cdot)$ on the set of actions $\mathcal{S}$, i.e.,
\begin{align}
\text{Pr}\{s[n]=s|q[1]=q_1,s[1]=s_1,\cdots,q[n]=q_n\}=p_{\delta_n(q_1,s_1,\cdots,q_n)}(s).
\end{align}

Define a transmission policy $\gamma=(\delta_1,\delta_2,\cdots)$, which is a sequence of decision rules. Define $\text{E}_{q_0}^{\gamma}\{\cdot\}$ as the notation of the expectation when policy $\gamma$ is applied and the initial state is $q_0$. Therefore the average power consumption under policy $\gamma$ 
\begin{align}
P_{\gamma}=\lim_{N\rightarrow \infty} \frac{1}{N} \text{E}_{q_0}^{\gamma} \left\{ \sum_{n=1}^N p[n] \right\}.
\label{eqn_P}
\end{align}

Let $D_{\gamma}$ denote the average delay under policy $\gamma$. According to Little's Law, the average queueing delay is the quotient of the average queue length divided by the average arrival rate, i.e.,
\begin{align}
D_{\gamma}=\frac{1}{\alpha A}\lim_{N\rightarrow \infty} \frac{1}{N} \text{E}_{q_0}^{\gamma} \left\{ \sum_{n=1}^N q[n] \right\}.
\label{eqn_D}
\end{align}

Therefore, policy $\gamma$ will determine $Z_{\gamma}=(P_{\gamma},D_{\gamma})$, which is a point in the delay-power plane. Define $\overline{Z_{\gamma}Z_{\gamma'}}$ as the line segment connecting $Z_{\gamma}$ and $Z_{\gamma'}$. Let $\Gamma$ denote the set of all feasible policies which can guarantee no overflow or underflow. Define $\mathcal{R}=\{Z_{\gamma} | \gamma \in \Gamma \}$ as the set of all feasible points in the delay-power plane. Intuitively, since the power consumption for each data packet increases if we want to transmit faster, there is a tradeoff between the average queueing delay and the average power consumption. Denote the optimal delay-power tradeoff curve $\mathcal{L}=\{(P,D)\in\mathcal{R}|\forall(P',D')\in\mathcal{R},\text{ either }P'\ge P\text{ or }D'\ge D\}$.

Since there are two costs in the MDP, by minimizing the average delay given an average power constraint, we obtain a CMDP problem.
\begin{subequations}
\label{eqn_original_optimization}
\begin{align}
\min\limits_{\gamma\in\Gamma}\quad & D_{\gamma}\\
\text{s.t.}\quad & P_{\gamma} \le P_{\text{th}}.
\end{align}
\end{subequations}
By varying the value of $P_{\text{th}}$, the optimal delay-power tradeoff curve $\mathcal{L}$ can be obtained. In the following, we show that optimizing over a simpler class of policies will minimize the objective in (\ref{eqn_original_optimization}).

\subsection{Reduction to Stationary Policies}
Here, we show that in order to solve our problem, it is enough to restrict our class of policies to a stationary class of policies. A stationary policy for an MDP means that the probability distribution to determine $s[n]$ is only a function of state $q[n]$, i.e. $\delta_n:\mathcal{Q} \rightarrow \mathcal{P}(\mathcal{S})$, and the decision rules for all timeslots are the same. For a CMDP, it is proven in \cite[Theorem 11.3]{altman1999constrained} that stationary policies are complete, which means stationary policies can achieve as good performance as any other policies. Therefore we only need to consider stationary policies in this problem.

Denote $f_{q,s}$ as the probability to transmit $s$ packet(s) when $q[n]=q$, i.e.
\begin{align}
f_{q,s}=\text{Pr}\{s[n]=s|q[n]=q\}.
\end{align}
Therefore we have
$\sum_{s=0}^{S}f_{q,s}=1$ for all $q=0,\cdots,Q$.
We guarantee the avoidance of overflow or underflow by setting $f_{q,s}=0$ if $q-s<0$ or $q-s>Q-A$. Denote $\boldsymbol{F}$ as a $(Q+1)\times(S+1)$ matrix whose element in the $(q+1)$th row and the $(s+1)$th column is $f_{q,s}$. Therefore matrix $\boldsymbol{F}$ can represent a decision rule, and moreover a stationary transmission policy. Denote $P_{\boldsymbol{F}}$ and $D_{\boldsymbol{F}}$ as the average power consumption and the average queueing delay under policy $\boldsymbol{F}$. Denote $\mathcal{F}$ as the set of all feasible stationary policies which can guarantee no overflow or underflow. Denote $\mathcal{F}_D$ as the set of all stationary and deterministic policies which can guarantee no overflow or underflow. Thus the optimization problem (\ref{eqn_original_optimization}) is equivalent to
\begin{subequations}
\label{eqn_stationary_optimization}
\begin{align}
\min\limits_{\boldsymbol{F}\in\mathcal{F}}\quad & D_{\boldsymbol{F}}\\
\text{s.t.}\quad & P_{\boldsymbol{F}} \le P_{\text{th}}.
\end{align}
\end{subequations}

\begin{figure*}[!t]
\centering
\begin{tikzpicture}[->, >=stealth', very thick, every node/.style={fill=white, font=\small}]
\node[state] (0) at (0,0) {$0$};
\node[state] (1) at (2.2,0) {$1$};
\node[state] (2) at (4.4,0) {$2$};
\node[state] (3) at (6.6,0) {$3$};
\node[state] (4) at (8.8,0) {$4$};
\node[state] (5) at (11,0) {$5$};
\node[state] (6) at (13.2,0) {$6$};
\node[state] (7) at (15.4,0) {$7$};

\path
(1) edge [bend right=15] node {$\lambda_{1,0}$} (0)
(2) edge [bend right=15] node {$\lambda_{2,1}$} (1)
(3) edge [bend right=15] node {$\lambda_{3,2}$} (2)
(4) edge [bend right=15] node {$\lambda_{4,3}$} (3)
(5) edge [bend right=15] node {$\lambda_{5,4}$} (4)
(6) edge [bend right=15] node {$\lambda_{6,5}$} (5)
(7) edge [bend right=15] node {$\lambda_{7,6}$} (6)

(2) edge [bend right=45] node {$\lambda_{2,0}$} (0)
(3) edge [bend right=45] node {$\lambda_{3,1}$} (1)
(4) edge [bend right=45] node {$\lambda_{4,2}$} (2)
(5) edge [bend right=45] node {$\lambda_{5,3}$} (3)
(6) edge [bend right=45] node {$\lambda_{6,4}$} (4)
(7) edge [bend right=45] node {$\lambda_{7,5}$} (5)

(3) edge [bend right=75] node {$\lambda_{3,0}$} (0)
(4) edge [bend right=75] node {$\lambda_{4,1}$} (1)
(5) edge [bend right=75] node {$\lambda_{5,2}$} (2)
(6) edge [bend right=75] node {$\lambda_{6,3}$} (3)
(7) edge [bend right=75] node {$\lambda_{7,4}$} (4)

(0) edge [bend right=15] node {$\lambda_{0,1}$} (1)
(1) edge [bend right=15] node {$\lambda_{1,2}$} (2)
(2) edge [bend right=15] node {$\lambda_{2,3}$} (3)
(3) edge [bend right=15] node {$\lambda_{3,4}$} (4)
(4) edge [bend right=15] node {$\lambda_{4,5}$} (5)
(5) edge [bend right=15] node {$\lambda_{5,6}$} (6)
(6) edge [bend right=15] node {$\lambda_{6,7}$} (7)

(0) edge [bend right=45] node {$\lambda_{0,2}$} (2)
(1) edge [bend right=45] node {$\lambda_{1,3}$} (3)
(2) edge [bend right=45] node {$\lambda_{2,4}$} (4)
(3) edge [bend right=45] node {$\lambda_{3,5}$} (5)
(4) edge [bend right=45] node {$\lambda_{4,6}$} (6)
(5) edge [bend right=45] node {$\lambda_{5,7}$} (7)
;
\end{tikzpicture}
\caption{Markov Chain of $q[n]$ ($Q=7$, $A=2$, $S=3$)}
\label{fig_markov}
\vspace{-0.8cm}
\end{figure*}
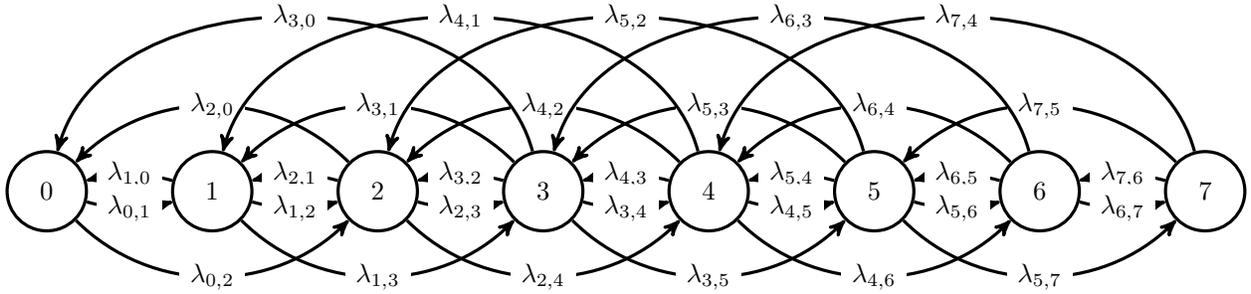

\subsection{Reduction to Unichains}
Given a stationary policy for a Markov Decision Process, there is an inherent Markov Reward Process (MRP) with $q[n]$ as the state variable. Denote $\lambda_{i,j}$ as the transition probability from state $i$ to state $j$. An example of the transition diagram is shown in \figurename~\ref{fig_markov}, where $\lambda_{i,i}$ for $i=0,\cdots,Q$ are omitted to keep the diagram legible.

The Markov chain could have more than one closed communication classes under certain transmission policies. For example, in the example in \figurename~\ref{fig_markov}, if we apply the scheduling policy $f_{0,0}=1$, $f_{1,0}=1$, $f_{2,2}=1$, $f_{3,2}=1$, $f_{4,2}=1$, $f_{5,2}=1$, $f_{6,2}=1$, $f_{7,2}=1$, and $f_{i,j}=0$ for all others, it can be seen that states 4, 5, 6 and 7 are transient, while states $\{0, 2\}$ and states $\{1, 3\}$ are two closed communication classes. Under this circumstances, the limiting probability distribution and the average cost are dependent on the initial state and the sample paths. However, the following theorem will show that we only need to study the cases with only one closed communication class.

\begin{theorem}
If the Markov chain generated by policy $\boldsymbol{F}$ has more than one closed communication class, named as $\mathcal{C}_1$, $\cdots$, $\mathcal{C}_L$, where $L>1$, then for all $1\le l \le L$, there exists a policy $\boldsymbol{F}_l$ such that the Markov chain generated by $\boldsymbol{F}_l$ has $\mathcal{C}_l$ as its only closed communication class. Moreover, the limiting distribution and the average cost of the Markov chain generated by $\boldsymbol{F}$ starting from state $c\in\mathcal{C}_l$ are the same as the limiting distribution and the average cost of the Markov chain generated by $\boldsymbol{F}_l$.
\label{theorem_closed_class}
\end{theorem}
\begin{IEEEproof}
See Appendix \ref{proof_closed_class}.
\end{IEEEproof}

Based on Theorem \ref{theorem_closed_class}, without loss of generality, we can focus on the Markov chains with only one closed communication class, which are called unichains. For a unichain, the initial state or the sample path won't affect the limiting distribution or the average cost, which means the parameter $q_0$ in $\text{E}_{q_0}^{\gamma}\{\cdot\}$ won't affect the value of the function.

As we will demonstrate in the following two sections, the optimal policies for the Constrained MDP problem and its Lagrangian relaxation problem are threshold-based. Here, we define that, a stationary policy $\boldsymbol{F}$ is threshold-based, if and only if there exist $(S+1)$ thresholds $0 \le q_{\boldsymbol{F}}(0) \le q_{\boldsymbol{F}}(1) \le \cdots \le q_{\boldsymbol{F}}(S) \le Q$, such that $f_{q,s}>0$ only when $q_{\boldsymbol{F}}(s-1) \le q \le q_{\boldsymbol{F}}(s)$ (we set $q_{\boldsymbol{F}}(-1)=-1$ for the inequality when $s=0$). It means that, under policy $\boldsymbol{F}$, when the queue state is larger than threshold $q_{\boldsymbol{F}}(s-1)$ and smaller than $q_{\boldsymbol{F}}(s)$, it transmits $s$ packet(s). When the queue state is equal to threshold $q_{\boldsymbol{F}}(s)$, it transmits $s$ or $(s+1)$ packet(s). Note that under this definition, probabilistic policies can also be threshold-based.

\section{Optimal Deterministic Threshold-Based Policy for the Lagrangian Relaxation Problem}
In (\ref{eqn_stationary_optimization}), we formulate the optimization problem as a Constrained MDP, which is difficult to solve in general. Therefore, we first study the Lagrangian relaxation of (\ref{eqn_stationary_optimization}) in this section, and prove that the optimal policy for the relaxation problem is deterministic and threshold-based. We will then use these results to show that the solution to the original non-relaxed CMDP problem is also of a threshold type.

Denote $\mu\ge 0$ as the Lagrange multiplier. Thus the Lagrangian relaxation of (\ref{eqn_stationary_optimization}) is
\begin{align}
\min\limits_{\boldsymbol{F}\in \mathcal{F}} & \lim_{N\rightarrow \infty} \frac{1}{N} \text{E}_{q_0}^{\boldsymbol{F}} \left\{ \frac{1}{\alpha A} \sum_{n=1}^N \left( q[n] + \alpha A \mu p[n] \right) \right\} - \mu P_{\text{th}}.
\label{eqn_lag_relax}
\end{align}
In (\ref{eqn_lag_relax}), the term $-\mu P_{\text{th}}$ is constant. Therefore, the Lagrangian relaxation problem is minimizing a constructed combined average cost $\left( q[n] + \alpha A \mu p[n] \right)$. This is an infinite-horizon Markov Decision Process with an average cost criterion, for which it is proven in \cite[Theorem 9.1.8]{puterman2014markov} that, there exists an optimal stationary deterministic policy. For a stationary deterministic policy $\boldsymbol{F}\in \mathcal{F}_D$, denote $s_{\boldsymbol{F}}(q)$ as the packet(s) to transmit when $q[n]=q$. In other words, we have $f_{q,s_{\boldsymbol{F}}(q)}=1$ for all $q$. Define $\eta=\alpha A \mu$. Therefore (\ref{eqn_lag_relax}) is equivalent to
\begin{align}
\min\limits_{\boldsymbol{F}\in \mathcal{F}_D}
\lim_{N\rightarrow \infty} \frac{1}{N} \text{E}_{q_0}^{\boldsymbol{F}} \left\{ \sum_{n=1}^N \left( q[n]+\eta P_{s_{\boldsymbol{F}}(q[n])}\right) \right\}.\label{eqn_opt_combined}
\end{align}
The optimal policy for (\ref{eqn_opt_combined}) has the following property.

\begin{theorem}
An optimal policy $\boldsymbol{F}$ for (\ref{eqn_opt_combined}) is threshold-based. That is to say, policy $\boldsymbol{F}$ should satisfy that $s_{\boldsymbol{F}}(q+1)-s_{\boldsymbol{F}}(q)=0$ or $1$ for all $0\le q<Q$.
\label{theorem_01}
\end{theorem}
\begin{IEEEproof}
For the simplicity of notations, in the proof we use $s(q)$ instead of $s_{\boldsymbol{F}}(q)$. Define
\begin{align}
h^{(m+1)}(q,s)=q+ \eta P_s+\alpha[h^{(m)}(q-s+A)-h^{(m)}(A)]+(1-\alpha)[h^{(m)}(q-s)-h^{(m)}(0)].
\end{align}

We will prove the theorem by applying a nested induction method to policy iteration algorithm for the Markov Decision Process. In Markov Decision Processes with an average cost, policy iteration algorithm can be applied to obtain the optimal scheduling policy, which is shown in Algorithm \ref{algo_policy_iteration}. In the algorithm, the function $h^{(m)}(q)$ converges to $h(q)$, which is called the potential function or bias function for the Markov Decision Process. The bias function can be interpreted as the expected total difference between the cost starting from a specific state and the stationary cost. The policy iteration algorithm can converge to the optimal solution in finite steps, which is proven in \cite[Theorem 8.6.6]{puterman2014markov} and \cite[Proposition 3.4]{bertsekas1995dynamic}.

The sketch of the proof is as follows. Because of the mechanism of the policy iteration algorithm, we can assign $h^{(0)}(q)$ as strictly convex in $q$. In Part I, we will demonstrate by induction that, for any $m$, if $h^{(m)}(q)$ is strictly convex in $q$, then $s^{(m+1)}(q)$ has the threshold-based property. In Part II, we will demonstrate that if $s^{(m+1)}(q)$ has the threshold-based property, then $h^{(m+1)}(q)$ is strictly convex in $q$. Therefore, by mathematical induction, we can prove the theorem.

\begin{algorithm}[t]
\caption{Policy Iteration Algorithm for Markov Decision Processes}
\begin{algorithmic}[1]
\State $m \gets 0$
\ForAll{$q$}
\State $h^{(0)}(q) \gets$ arbitrary value // Initialization
\EndFor
\Repeat
\ForAll{$q$}
\State $s^{(m+1)}(q) \gets \arg\min_s\{h^{(m+1)}(q,s)\}$ // Policy Improvement
\EndFor
\ForAll{$q$}
\State $h^{(m+1)}(q) \gets h^{(m+1)}(q,s^{(m+1)}(q))$ // Policy Evaluation
\EndFor
\State $m \gets m+1$
\Until{$s^{(m)}(q)=s^{(m-1)}(q)$ holds for all $q$}
\State $s(q) \gets s^{(m)}(q)$ for all $q$
\end{algorithmic}
\label{algo_policy_iteration}
\end{algorithm}

\textbf{Part I. Convexity of $h^{(m)}(q)$ in $q$ $\rightarrow$ threshold-based property of $s^{(m+1)}(q)$}

Assume $h^{(m)}(q)$ is strictly convex in $q$. In this part, we are going to prove $s^{(m+1)}(q)$ has the threshold-based property.
\begin{enumerate}
\item Because of the requirements of a feasible policy, we have $s^{(m+1)}(0)=0$, and $s^{(m+1)}(1)=0$ or $1$. Therefore $s^{(m+1)}(q+1)-s^{(m+1)}(q)=0$ or $1$ when $q=0$.
\item We define $s_1=s^{(m+1)}(q_1)$ for a specific $q_1$. From the Policy Improvement step in the policy iteration algorithm, we have the following inequalities:
\begin{align}
h^{(m+1)}(q_1,s_1) \le & h^{(m+1)}(q_1,s_1-\delta), \forall 0\le \delta \le s_1,\label{eqn_t1_left}\\
h^{(m+1)}(q_1,s_1) \le & h^{(m+1)}(q_1,s_1+\delta), \forall 0\le \delta \le S-s_1.\label{eqn_t1_right}
\end{align}

Since $h^{(m)}(q)$ is strictly convex in $q$,
\begin{align}
& h^{(m)}(q_1+1-s_1+A)-h^{(m)}(q_1-s_1+A) \nonumber\\
< & h^{(m)}(q_1+1-(s_1-\delta)+A) -h^{(m)}(q_1-(s_1-\delta)+A),\label{eqn_h_convex_1}\\
& h^{(m)}(q_1+1-s_1)-h^{(m)}(q_1-s_1) \nonumber\\
< & h^{(m)}(q_1+1-(s_1-\delta))-h^{(m)}(q_1-(s_1-\delta)).\label{eqn_h_convex_2}
\end{align}

Since $P_s$ is strictly convex, 
\begin{align}
P_{s_1+1}-P_{s_1}<P_{s_1+1+\delta}-P_{s_1+\delta}.\label{eqn_Ps_convex}
\end{align}
By computing (\ref{eqn_t1_left})$+\alpha\times$(\ref{eqn_h_convex_1})$+(1-\alpha)\times$(\ref{eqn_h_convex_2}), we have
\begin{align}
h^{(m+1)}(q_1+1,s_1) < h^{(m+1)}(q_1+1,s_1-\delta), \forall 0 \le \delta \le s_1.\label{eqn_t1+1_left}
\end{align}
By computing (\ref{eqn_t1_right}) and (\ref{eqn_Ps_convex}), we have
\begin{align}
h^{(m+1)}(q_1+1,s_1+1) < h^{(m+1)}(q_1+1,s_1+1+\delta), \forall 0 \le \delta \le S-s_1-1.\label{eqn_t1+1_right}
\end{align}

From (\ref{eqn_t1+1_left}) and (\ref{eqn_t1+1_right}), we can see that $s^{(m+1)}(q_1+1)$ can only be $s_1$ or $s_1+1$. In other words, we have $s^{(m+1)}(q_1+1)-s^{(m+1)}(q_1)=0$ or $1$.
\end{enumerate}

From above, by mathematical induction, we can have that $s^{(m+1)}(q)$ has the threshold-based property.

\textbf{Part II. Threshold-based property of $s^{(m+1)}(q)$ $\rightarrow$ convexity of $h^{(m+1)}(q)$ in $q$}

Assume $s^{(m+1)}(q)$ has the threshold-based property. We still use the same notation as in the previous part that $s_1=s^{(m+1)}(q_1)$ for a specific $q_1$, and $s^{(m+1)}(q_1+1)=s_1$ or $s_1+1$.
\begin{enumerate}
\item If $s^{(m+1)}(q_1+1)=s_1$,
\begin{align}
& h^{(m+1)}(q_1+1)-h^{(m+1)}(q_1) \le h^{(m+1)}(q_1+1, s_1+1)-h^{(m+1)}(q_1,s_1)\\
= &  (q_1+1)+ \eta P_{s_1+1}+\alpha[h^{(m)}((q_1+1)-(s_1+1)+A)-h^{(m)}(A)]\nonumber\\
& +(1-\alpha)[h^{(m)}((q_1+1)-(s_1+1))-h^{(m)}(0)]\nonumber\\
& -[ q_1+ \eta P_{s_1}+\alpha[h^{(m)}(q_1-s_1+A)-h^{(m)}(A)]\nonumber\\
& +(1-\alpha)[h^{(m)}(q_1-s_1)-h^{(m)}(0)]]\\
= & 1+\eta (P_{s_1+1}-P_{s_1}).
\end{align}
On the other hand,
\begin{align}
& h^{(m+1)}(q_1+1)-h^{(m+1)}(q_1) > h^{(m+1)}(q_1+1,s_1)-h^{(m+1)}(q_1,s_1-1)\\
=& (q_1+1)+\eta P_{s_1}+\alpha[h^{(m)}((q_1+1)-s_1+A)-h^{(m)}(A)]\nonumber\\
& +(1-\alpha)[h^{(m)}((q_1+1)-s_1)-h^{(m)}(0)]\nonumber\\
& -[q_1+ \eta P_{s_1-1}+\alpha[h^{(m)}(q_1-(s_1-1)+A)-h^{(m)}(A)]\nonumber\\
& +(1-\alpha)[h^{(m)}(q_1-(s_1-1))-h^{(m)}(0)]]\\
=& 1+\eta (P_{s_1}-P_{s_1-1}).
\end{align}
\item If $s^{(m+1)}(q_1+1)=s_1+1$,
\begin{align}
& h^{(m+1)}(q_1+1)-h^{(m+1)}(q_1)\nonumber\\
=& (q_1+1)+ \eta P_{s_1+1}+\alpha[h^{(m)}((q_1+1)-(s_1+1)+A)-h^{(m)}(A)]\nonumber\\
& +(1-\alpha)[h^{(m)}((q_1+1)-(s_1+1))-h^{(m)}(0)]\nonumber\\
& -[q_1+ \eta P_{s_1}+\alpha[h^{(m)}(q_1-s_1+A)-h^{(m)}(A)]\nonumber\\
& +(1-\alpha)[h^{(m)}(q_1-s_1)-h^{(m)}(0)]]\\
=&1+ \eta (P_{s_1+1}-P_{s_1}).
\end{align}
\end{enumerate}

To conclude, $1+\eta(P_{s_1}-P_{s_1-1}) < h^{(m+1)}(q_1+1)-h^{(m+1)}(q_1) \le 1+\eta(P_{s_1+1}-P_{s_1})$ holds for any specific $q_1$. Therefore $h^{(m+1)}(q+1)-h^{(m+1)}(q)$ is strictly increasing, which means $h^{(m+1)}(q)$ is strictly convex in $q$.

Based on the assumption for initial $h^{(0)}(q)$ and the derivations in Part I and II, by mathematical induction, we can prove that $s^{(m)}(q)$ has the threshold-based property for all $m \ge 1$. Since $s^{(m)}(q)$ will converge to the optimal policy $s(q)$ in finite steps, the optimal policy $s(q)$ has the threshold-based property.
\end{IEEEproof}

Theorem \ref{theorem_01} indicates a very intuitive conclusion that more data should be transmitted if the queue is longer. More specifically speaking, for an optimal deterministic policy $\boldsymbol{F}$, there exists $(S+1)$ thresholds $q_{\boldsymbol{F}}(0) \le q_{\boldsymbol{F}}(1) \le \cdots \le q_{\boldsymbol{F}}(S)$, such that
\begin{align}
\begin{cases}
f_{q,s}=1 \qquad q_{\boldsymbol{F}}(s-1)<q\le q_{\boldsymbol{F}}(s), s=0,\cdots,S\\
f_{q,s}=0 \qquad \text{else}
\end{cases}
\label{eqn_deterministic_threshold}
\end{align}
where $q_{\boldsymbol{F}}(-1)=-1$. The form of the optimal policy satisfies our definition of threshold-based policy in Section II.

Moreover, we can have the following two corollaries.

\begin{corollary}
Under any optimal threshold-based policy $\boldsymbol{F}$, there will be no transmission only when $q[n]=0$. In other words, threshold $q_{\boldsymbol{F}}(0)=0$.
\end{corollary}
\begin{IEEEproof}
This is an intuitive result, because every data packet will be transmitted sooner or later, which costs at least $P_1$ power, thus not transmitting when there are backlogs is just a waste of time. The following is its rigorous proof.

If there exists an optimal threshold-based policy $\boldsymbol{F}\in \mathcal{F}_D$ where $s_{\boldsymbol{F}}(q_1)=0$, $q_1>0$. Since $s_{\boldsymbol{F}}(q)$ has the threshold-based property, we have $s_{\boldsymbol{F}}(1)=0$. Construct a policy $\boldsymbol{F}'$ where $s_{\boldsymbol{F}'}(q)=s_{\boldsymbol{F}}(q+1)$ for $0\le q<Q$ and $s_{\boldsymbol{F}'}(Q)=A$. It can be seen that $\boldsymbol{F}'\in \mathcal{F}_D$. State $0$ is a transient state under policy $\boldsymbol{F}$, and state $Q$ is a transient state under policy $\boldsymbol{F}'$. States $1,\cdots,Q$ under policy $\boldsymbol{F}$ and states $0,\cdots,Q-1$ under policy $\boldsymbol{F}'$ have the exactly same state transition, except that the states for $\boldsymbol{F}'$ are 1 smaller than the states for $\boldsymbol{F}$. Therefore the average power consumption under two policies is the same and the average queue length for $\boldsymbol{F}'$ is 1 smaller, which means the average delay for $\boldsymbol{F}'$ is strictly smaller. Therefore $\boldsymbol{F}$ is not an optimal policy, which conflicts with the assumption. Hence the optimal threshold-based policy should have that there will be no transmissions only when $q[n]=0$.
\end{IEEEproof}

\begin{corollary}
For an optimal threshold-based policy $\boldsymbol{F}$, there is no need to transmit more than $A$ packets. In other words, threshold $q_{\boldsymbol{F}}(A)=q_{\boldsymbol{F}}(A+1)=\cdots=q_{\boldsymbol{F}}(S)=Q$.
\end{corollary}
\begin{IEEEproof}
If there exists an optimal threshold-based policy $\boldsymbol{F}\in\mathcal{F}_D$ where $q_1$ is the smallest state such that $s_{\boldsymbol{F}}(q_1)>A$. Since $s_{\boldsymbol{F}}(q)$ has the threshold-based property, for all $q\ge q_1$, we have $s(q)>A$. Also, for all $q<q_1$, we have $s(q)\le A$. Construct a policy $\boldsymbol{F}'$ where $s_{\boldsymbol{F}'}(q)=s_{\boldsymbol{F}}(q)$ for $q<q_1$ and $s_{\boldsymbol{F}'}(q)=A$ for $q\ge q_1$. It can be seen that $\boldsymbol{F}'\in\mathcal{F}_D$. Since $q_1,\cdots,Q$ are transient states under both policies, and the transmission is exactly the same for both policies, policy $\boldsymbol{F}'$ has the same performance as policy $\boldsymbol{F}$. Therefore, for an optimal threshold-based policy, there is no need to transmit more than $A$ packets.
\end{IEEEproof}

\section{Optimal Threshold-Based Policy for the CMDP}
In Section III, we prove that the optimal policy to minimize the combined cost is deterministic and threshold-based. We will now prove that the solution to the overall CMDP problem also takes on a threshold form. We first conduct steady-state analysis for the Markov Decision Process, discover that the feasible average delay and power region is a convex polygon and the optimal delay-power tradeoff curve is piecewise linear, whose neighbouring vertices are obtained by deterministic policies which take different actions in only one state. Based on this, the optimal threshold-based policy obtained in Section III will be shown to correspond to the vertices of the piecewise linear curve. Therefore, the optimal policy for the CMDP problem, which is the convex combination of two deterministic threshold-based policies, will be proven to also take a threshold form. Then, we will provide an efficient algorithm to obtain the optimal delay-power tradeoff curve, and a Linear Programming will be formulated to confirm our results.

Based on Theorem \ref{theorem_closed_class}, without loss of generality, we can focus on unichains, in which case the steady-state probability distribution exists. Denote $\pi_{\boldsymbol{F}}(q)$ as the steady-state probability for state $q$ when applying policy $\boldsymbol{F}$. Denote $\boldsymbol{\pi}_{\boldsymbol{F}}=[\pi_{\boldsymbol{F}}(0),\cdots,\pi_{\boldsymbol{F}}(Q)]^T$. Denote $\boldsymbol{\Lambda}_{\boldsymbol{F}}$ as a $(Q+1)\times(Q+1)$ matrix whose element in the $(i+1)$th column and the $(j+1)$th row is $\lambda_{i,j}$, which is determined by policy $\boldsymbol{F}$. Denote $\boldsymbol{I}$ as the identity matrix. Denote $\boldsymbol{1}=[1,\cdots,1]^T$, and $\boldsymbol{0}=[0,\cdots,0]^T$. We won't specify the size of $\boldsymbol{I}$, $\boldsymbol{1}$ or $\boldsymbol{0}$ if there is no ambiguity. Denote $\boldsymbol{G}_{\boldsymbol{F}}=\boldsymbol{\Lambda}_{\boldsymbol{F}}-\boldsymbol{I}$. Denote
$\boldsymbol{H}_{\boldsymbol{F}}=\left[
\begin{array}{c}
\boldsymbol{1}^T\\
\boldsymbol{G}_{\boldsymbol{F}}(0:(Q-1),:)
\end{array}
\right]
$ and
$\boldsymbol{c}=\left[
\begin{array}{c}
1\\\boldsymbol{0}
\end{array}
\right]$.

From the definition of the steady-state distribution, we have $\boldsymbol{G}_{\boldsymbol{F}}\boldsymbol{\pi}_{\boldsymbol{F}}=\boldsymbol{0}$ and $\boldsymbol{1}^T\boldsymbol{\pi}_{\boldsymbol{F}}=1$. For a unichain, the rank of $\boldsymbol{G}_{\boldsymbol{F}}$ is $Q$. Therefore, we have $\boldsymbol{H}_{\boldsymbol{F}}$ is invertible and
\begin{align}
\boldsymbol{\pi}_{\boldsymbol{F}}=\boldsymbol{H}_{\boldsymbol{F}}^{-1}\boldsymbol{c}.
\label{pi-H}
\end{align}

We can express the average power consumption $P_{\boldsymbol{F}}$ and the average delay $D_{\boldsymbol{F}}$ using the steady-state probability distribution. For state $q$, transmitting $s$ packet(s) will cost $P_s$ with probability $f_{q,s}$. Denote $\boldsymbol{p}_{\boldsymbol{F}}=[\sum_{s=0}^S P_s f_{0,s},\cdots,\sum_{s=0}^S P_s f_{Q,s}]^T$, which is a function of $\boldsymbol{F}$, thus the average power consumption
\begin{align}
P_{\boldsymbol{F}}=\sum_{q=0}^{Q} \pi_{\boldsymbol{F}}(q) \sum_{s=0}^S P_s f_{q,s}=\boldsymbol{p}_{\boldsymbol{F}}^T\boldsymbol{\pi}_{\boldsymbol{F}}.
\label{PwithPi}
\end{align}
Similarly, denote $\boldsymbol{d}=[0,1,\cdots,Q]^T$, thus the average delay under policy $\boldsymbol{F}$
\begin{align}
D_{\boldsymbol{F}}=\frac{1}{\alpha A}\sum_{q=0}^{Q} q \pi_{\boldsymbol{F}}(q)=\frac{1}{\alpha A}\boldsymbol{d}^T\boldsymbol{\pi}_{\boldsymbol{F}}.
\label{DwithPi}
\end{align}

\subsection{Partially Linear Property of Scheduling Policies}
The mapping from $\boldsymbol{F}$ to $Z_{\boldsymbol{F}}=(P_{\boldsymbol{F}},D_{\boldsymbol{F}})$ has a partially linear property shown in the following lemma.

\begin{lemma}
$\boldsymbol{F}$ and $\boldsymbol{F}'$ are two scheduling policies that are different only when $q[n]=q$, i.e. the two matrices are different only in the $(q+1)$th row. Denote $\boldsymbol{F}''=(1-\epsilon)\boldsymbol{F}+\epsilon\boldsymbol{F}'$ where $0\le \epsilon\le 1$. Then\newline
1) There exists a certain $0\le \epsilon'\le 1$ so that $P_{\boldsymbol{F}''}=(1-\epsilon')P_{\boldsymbol{F}}+\epsilon' P_{\boldsymbol{F}'}$ and $D_{\boldsymbol{F}''}=(1-\epsilon')D_{\boldsymbol{F}}+\epsilon' D_{\boldsymbol{F}'}$. Moreover, it should hold that $\epsilon'$ is a continuous nondecreasing function of $\epsilon$.\newline
2) When $\epsilon$ changes from 0 to 1, point $Z_{\boldsymbol{F}''}$ moves on the line segment $\overline{Z_{\boldsymbol{F}}Z_{\boldsymbol{F}'}}$ from $Z_{\boldsymbol{F}}$ to $Z_{\boldsymbol{F}'}$.
\label{lemma_linearcombination}
\end{lemma}
\begin{IEEEproof}
See Appendix \ref{proof_linearcombination}.
\end{IEEEproof}

Lemma \ref{lemma_linearcombination} indicates that the convex combination of scheduling policies which take different actions in only one state will induce the convex combination of points in the delay-power plane. Furthermore, we can have the following two further results.
\begin{theorem}
The set of all feasible points in the delay-power plane, $\mathcal{R}$, is a convex polygon whose vertices are all obtained by deterministic scheduling policies. Moreover, the policies corresponding to neighbouring vertices of $\mathcal{R}$ take different actions in only one state.
\label{theorem_region=polygon}
\end{theorem}
\begin{IEEEproof}
See Appendix \ref{proof_region=polygon}.
\end{IEEEproof}

\begin{corollary}
The optimal delay-power tradeoff curve $\mathcal{L}$ is piecewise linear, decreasing, and convex. The vertices of the curve are obtained by deterministic scheduling policies. Moreover, the policies corresponding to neighbouring vertices of $\mathcal{L}$ take different actions in only one state.
\label{corollary_piecewiselinear}
\end{corollary}
\begin{IEEEproof}
See Appendix \ref{proof_piecewiselinear}.
\end{IEEEproof}

\subsection{Optimal Threshold-Based Policy for the CMDP}
\begin{figure}[t]
\centering
\includegraphics[width=0.5\columnwidth]{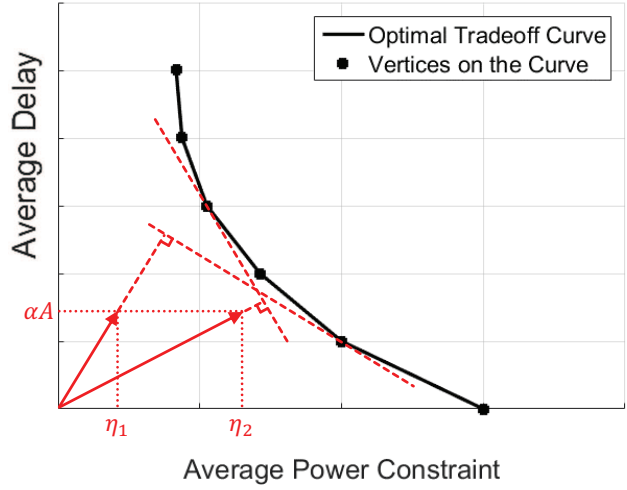}
\caption{The minimum inner product of points on $\mathcal{L}$ and the weighted vector can always be obtained by vertices of $\mathcal{L}$}
\label{fig_weightedsum_piecewiselinear}
\vspace{-0.8cm}
\end{figure}

In the last section, we prove in Theorem \ref{theorem_01} that the optimal policy for the combined cost is deterministic and threshold-based. Based on the steady-state analysis, the objective function in the unconstrained MDP problem (\ref{eqn_opt_combined})
\begin{align}
\lim_{N\rightarrow \infty} \frac{1}{N} \text{E}_{q_0}^{\boldsymbol{F}} \left\{ \sum_{n=1}^N \left( q[n]+\eta P_{s_{\boldsymbol{F}}(q[n])}\right) \right\}=\alpha A D_{\boldsymbol{F}}+\eta P_{\boldsymbol{F}}=\langle(\eta, \alpha A),Z_{\boldsymbol{F}}\rangle
\end{align}
can be seen as the inner product of vector $(\eta, \alpha A)$ and $Z_{\boldsymbol{F}}$. Since $\mathcal{R}$ is a convex polygon, the corresponding $Z_{\boldsymbol{F}}$ minimizing the inner product will be obtained by vertices of $\mathcal{L}$, as demonstrated in \figurename~\ref{fig_weightedsum_piecewiselinear}. Since the conclusion in Theorem \ref{theorem_01} holds for any $\eta$, the vertices of the optimal tradeoff curve are all obtained by optimal policies for the relaxed problem, which are deterministic and threshold-based. Moreover, from Corollary \ref{corollary_piecewiselinear}, the neighbouring vertices of $\mathcal{L}$ are obtained by policies which take different actions in only one state. Therefore, we have the following theorem.

\begin{theorem}
Given an average power constraint, the scheduling policy $\boldsymbol{F}$ to minimize the average delay takes the following form that, there exists $(S+1)$ thresholds $q_{\boldsymbol{F}}(0) \le q_{\boldsymbol{F}}(1) \le \cdots \le q_{\boldsymbol{F}}(S)$, one of which we denote as $q_{\boldsymbol{F}}(s^*)$, such that
\begin{align}
\left\{
\begin{array}{@{}ll}
f_{q,s}=1 & q_{\boldsymbol{F}}(s-1)<q\le q_{\boldsymbol{F}}(s),s\neq s^*\\
f_{q,s^*}=1\qquad\qquad & q_{\boldsymbol{F}}(s^*-1)<q< q_{\boldsymbol{F}}(s^*)\\
\multicolumn{2}{@{}l}{f_{q_{\boldsymbol{F}}(s^*),s^*}+f_{q_{\boldsymbol{F}}(s^*),s^*+1}=1} \\
f_{q,s}=0 & \text{else}
\end{array}
\right.
\label{eqn_threshold}
\end{align}
where $q_{\boldsymbol{F}}(-1)=-1$.
\label{theorem_threshold}
\end{theorem}
\begin{IEEEproof}
According to Corollary \ref{corollary_piecewiselinear}, the policies corresponding to neighbouring vertices of $\mathcal{L}$ are deterministic and take different actions in only one state. In other words, according to (\ref{eqn_deterministic_threshold}), the thresholds for $\boldsymbol{F}$ and $\boldsymbol{F}'$ are all the same except one of their thresholds are different by 1. Denote the thresholds for $\boldsymbol{F}$ as $q_{\boldsymbol{F}}(0),q_{\boldsymbol{F}}(1),\cdots,q_{\boldsymbol{F}}(s^*),\cdots,q_{\boldsymbol{F}}(S)$, and the thresholds for $\boldsymbol{F}'$ as $q_{\boldsymbol{F}}(0),q_{\boldsymbol{F}}(1),\cdots,q_{\boldsymbol{F}}(s^*)-1,\cdots,q_{\boldsymbol{F}}(S)$, which means the two policies are different only in state $q_{\boldsymbol{F}}(s^*)$. Since the policy to obtain a point on $\overline{Z_{\boldsymbol{F}}Z_{\boldsymbol{F}'}}$ is the convex combination of $\boldsymbol{F}$ and $\boldsymbol{F}'$, it should have the form shown in (\ref{eqn_threshold}).
\end{IEEEproof}

According to Theorem \ref{theorem_threshold}, policies corresponding to the points between vertices of the optimal tradeoff curve, as the mixture of two deterministic threshold-based policies different only in one state, also satisfies our definition of threshold-based policy in Section II. When $q_{\boldsymbol{F}}(s-1)<q[n]<q_{\boldsymbol{F}}(s)$, we transmit $s$ packet(s). Any optimal scheduling policy $\boldsymbol{F}$ has at most two decimal elements $f_{q_{\boldsymbol{F}}(s^*),s^*}$ and $f_{q_{\boldsymbol{F}}(s^*),s^*+1}$, while the other elements are either 0 or 1.

\subsection{Algorithm to Obtain the Optimal Tradeoff Curve}
Here, we propose Algorithm \ref{algo_curve} to efficiently obtain the optimal delay-power tradeoff curve. This algorithm is based on the results that the optimal delay-power tradeoff curve is piecewise linear, whose vertices are obtained by deterministic threshold-based policies, and policies corresponding to two adjacent vertices take different actions in only one state. With the optimal tradeoff curve obtained, the minimum delay given a specific power constraint can also be obtained.

\begin{algorithm}[t]
\caption{Constructing the Optimal Delay-Power Tradeoff}
\begin{minipage}[t]{0.5\textwidth}
\small
\begin{algorithmic}[1]
\State Construct $\boldsymbol{F}$ whose thresholds $q_{\boldsymbol{F}}(s)=s$ for $s < A$ and $q_{\boldsymbol{F}}(s)=Q$ for $s \ge A$
\State Calculate $D_{\boldsymbol{F}}$ and $P_{\boldsymbol{F}}$
\State $\mathcal{F}_c \gets [\boldsymbol{F}]$, $D_c \gets D_{\boldsymbol{F}}$, $P_c \gets D_{\boldsymbol{F}}$
\While{$\mathcal{F}_c \neq \emptyset$}
\State $\mathcal{F}_p \gets \mathcal{F}_c$, $D_p \gets D_c$, $P_p \gets D_c$
\State $\mathcal{F}_c \gets \emptyset$, $slope \gets +\infty$
\While{$\mathcal{F}_p \neq \emptyset$}
\State{$\boldsymbol{F}$=$\mathcal{F}_p$.pop(0)}
\ForAll{$0 < s^* < A$}
\State
\begin{tabular}{@{}l}
Construct $\boldsymbol{F}'$ where $q_{\boldsymbol{F}'}(s^*)=q_{\boldsymbol{F}}(s^*)+1$ \\ and $q_{\boldsymbol{F}'}(s)=q_{\boldsymbol{F}}(s)$ for $s\neq s^*$
\end{tabular}
\State NewPolicyProbing()
\State // Probing all possible candidates
\EndFor
\EndWhile
\State
\begin{tabular}{@{}l}
Draw the line segment connecting $(P_p,D_p)$ and \\ $(P_c,D_c)$
\end{tabular}
\EndWhile
\Statex
\end{algorithmic}
\end{minipage}
\begin{minipage}[t]{0.5\textwidth}
\small
\begin{algorithmic}[1]
\Procedure{NewPolicyProbing}{ }
\If{$\boldsymbol{F}'$ is feasible and threshold-based}
\State Calculate $D_{\boldsymbol{F}'}$ and $P_{\boldsymbol{F}'}$
\If{$D_{\boldsymbol{F}'} = D_p$ and $P_{\boldsymbol{F}'} = P_p$}
\State $\mathcal{F}_p$.append$(\boldsymbol{F}')$
\State // $Z_{\boldsymbol{F}'}$ coincides with $Z_p$
\ElsIf{$D_{\boldsymbol{F}'} \ge D_p$ and $P_{\boldsymbol{F}'} < P_p$}
\If{$\frac{D_{\boldsymbol{F}'}-D_p}{P_p-P_{\boldsymbol{F}'}}<slope$}
\State $\mathcal{F}_c \gets [\boldsymbol{F}']$, $slope \gets \frac{D_{\boldsymbol{F}'}-D_p}{P_p-P_{\boldsymbol{F}'}}$
\State $D_c \gets D_{\boldsymbol{F}'}$, $P_c \gets P_{\boldsymbol{F}'}$
\State // $Z_{\boldsymbol{F}'}$ has the best performance
\ElsIf{$\frac{D_{\boldsymbol{F}'}-D_p}{P_p-P_{\boldsymbol{F}'}}=slope$}
\If{$P_{\boldsymbol{F}'}=P_c$}
\State $\mathcal{F}_c$.append$(\boldsymbol{F}')$
\State
\begin{tabular}{@{}l}
// $Z_{\boldsymbol{F}'}$ has the same performance\\
~~ as the current best candidate(s)
\end{tabular}
\ElsIf{$P_{\boldsymbol{F}'}>P_c$}
\State $\mathcal{F}_c \gets [\boldsymbol{F}']$, $slope \gets \frac{D_{\boldsymbol{F}'}-D_p}{P_p-P_{\boldsymbol{F}'}}$
\State $D_c \gets D_{\boldsymbol{F}'}$, $P_c \gets P_{\boldsymbol{F}'}$
\State
\begin{tabular}{@{}l}
// $Z_{\boldsymbol{F}'}$ has the same slope\\
~~ as the current best candidate(s)\\
~~ but is closer to $Z_p$
\end{tabular}
\EndIf
\EndIf
\EndIf
\EndIf
\EndProcedure
\end{algorithmic}
\end{minipage}
\label{algo_curve}
\end{algorithm}

The basic idea of the algorithm is, we start from the bottom-right vertex of the optimal tradeoff curve, whose corresponding policy is to transmit as much as possible. Then we enumerate all the candidates for the next vertex of the curve, based on the conclusion that policies corresponding to adjacent vertices take different actions in only one threshold. The next vertex will be determined by the policy candidate whose connecting line with the current vertex has the minimum absolute slope and the minimum length. Note that a vertex can be obtained by more than one policy, therefore we use lists $\mathcal{F}_p$ and $\mathcal{F}_c$ to restore all policies corresponding to the previous and the current vertices. When conducting the complexity analysis, we assume the situation where a vertex is obtained by multiple policies rarely happens. Since one of the thresholds of the policy will be decreased by 1 during each iteration, the maximum iteration number is $AQ$. Within each iteration, there are $A$ thresholds to try. In each trial, the most time consuming operation, the matrix inversion, costs $\boldsymbol{O}(Q^3)$. Therefore the complexity of the algorithm is $\boldsymbol{O}(A^2Q^4)$.

\subsection{Linear Programming Formulation}
In the following, we demonstrate that, the CMDP problem can also be formulated as a Linear Programming, which can be solved for a certain power constraint. We compare Algorithm \ref{algo_curve} and Linear Programming, and demonstrate that our algorithm is superior to the Linear Programming based approach. In the next section, we will use Linear Programming to confirm the properties of the optimal tradeoff curve and the algorithm we have demonstrated.

Based on the steady-state analysis (\ref{pi-H}) (\ref{PwithPi}) and (\ref{DwithPi}), the optimization problem (\ref{eqn_stationary_optimization}) can be transformed into
\begin{subequations}
\label{eqn_opt_steady}
\begin{align}
\min\limits_{\boldsymbol{F},\boldsymbol{\pi}}\quad
& \frac{1}{\alpha A}\boldsymbol{d}^T\boldsymbol{\pi}_{\boldsymbol{F}}\\
\text{s.t.}\quad
& \boldsymbol{p}_{\boldsymbol{F}}^T\boldsymbol{\pi}_{\boldsymbol{F}} \le P_{\text{th}}\\
& \boldsymbol{H}_{\boldsymbol{F}}\boldsymbol{\pi}_{\boldsymbol{F}}=\boldsymbol{c}\\
& \boldsymbol{\pi}_{\boldsymbol{F}} \succeq \boldsymbol{0}\\
& f_{q,s}=0 \qquad \forall q-s<0 \text{ or } q-s>Q-A
\end{align}
\end{subequations}
where $\boldsymbol{\pi}_{\boldsymbol{F}} \succeq \boldsymbol{0}$ means $\boldsymbol{\pi}$ is componentwise nonnegative.

Define $x_{q,s}=\pi(q) f_{q,s}$. By substituting the variables in (\ref{eqn_opt_steady}), the optimization problem can be transformed into
\begin{subequations}
\label{eqn_opt_lp}
\begin{align}
\min\quad
& \frac{1}{\alpha A}\sum_{q=0}^{Q} q \sum_{s=0}^S x_{q,s}\\
\text{s.t.}\quad
& \sum_{q=0}^{Q} \sum_{s=0}^S P_s x_{q,s} \le P_{\text{th}}\\
& \sum_{l=\max\{0,q-A\}}^{q-1}\sum_{s=0}^{l+A-q}\alpha x_{l,s}=\sum_{r=q}^{\min\{q+S-1,Q\}}\sum_{s=r-q+A+1}^S x_{r,s}\nonumber\\
&+\sum_{r=q}^{\min\{q+S-1,Q\}}\sum_{s=r-q+1}^{r-q+A} (1-\alpha) x_{r,s} \quad q=1,\cdots,Q\\
& \sum_{q=0}^{Q} \sum_{s=0}^S x_{q,s}=1\\
& x_{q,s}=0 \qquad \forall q-s<0 \text{ or } q-s>Q-A\\
& x_{q,s}\ge 0 \qquad \forall 0 \le q-s \le Q-A.
\end{align}
\end{subequations}

It can be observed that this is a Linear Programming problem. Given a feasible solution to (\ref{eqn_opt_steady}), $\boldsymbol{F}$ and $\boldsymbol{\pi}$, it can be checked that $x_{q,s}=\pi(q) f_{q,s}$ for all $q$ and $s$ is a feasible solution to (\ref{eqn_opt_lp}) with the same objective value. On the other hand, given a feasible solution to (\ref{eqn_opt_lp}), $x_{q,s}$ for all $q$ and $s$, it can be proven that $\pi(q)=\sum_{s=0}^{S}x_{q,s}$ and $f_{q,s}=\begin{cases}
\frac{x_{q,s}}{\pi(q)} & \pi(q)>0\\
1 & \pi(q)=0, s=\min\{q,S\}\\
0 & \pi(q)=0, s\neq \min\{q,S\}
\end{cases}$ is a feasible solution to (\ref{eqn_opt_steady}) with the same objective value. This means Linear Programming (\ref{eqn_opt_lp}) is equivalent with (\ref{eqn_opt_steady}), thus also equivalent with (\ref{eqn_opt_steady}).

If we apply the ellipsoid algorithm to solve (\ref{eqn_opt_lp}), the computational complexity is $\boldsymbol{O}(A^4Q^4)$. It means that, applying Linear Programming to obtain one point on the optimal tradeoff curve consumes more computation than obtaining the entire curve with Algorithm \ref{algo_curve}. Moreover, when the average energy constraint is dynamically changing, Linear Programming needs to be solved for each constraint, while the constructed power-delay trade-off from Algorithm \ref{algo_curve} can adapt to the changed constraint instantly. This demonstrates the inherent advantage in using the revealed properties of the optimal tradeoff curve and the optimal policies.

\section{Numerical Results}
In this section, we validate our theoretical results by conducting numerical computation and simulations. The convex feasible delay-power region and the generated polygons will be demonstrated in a small-scale example. The delay-power tradeoff curves will be obtained in a practical scenario. It will be confirmed that Algorithm \ref{algo_curve} can obtain the optimal delay-power tradeoff curve for both cases.

\begin{figure}[!tb] 
\begin{minipage}[b]{0.5\textwidth} 
\centering 
\includegraphics[width=1\columnwidth]{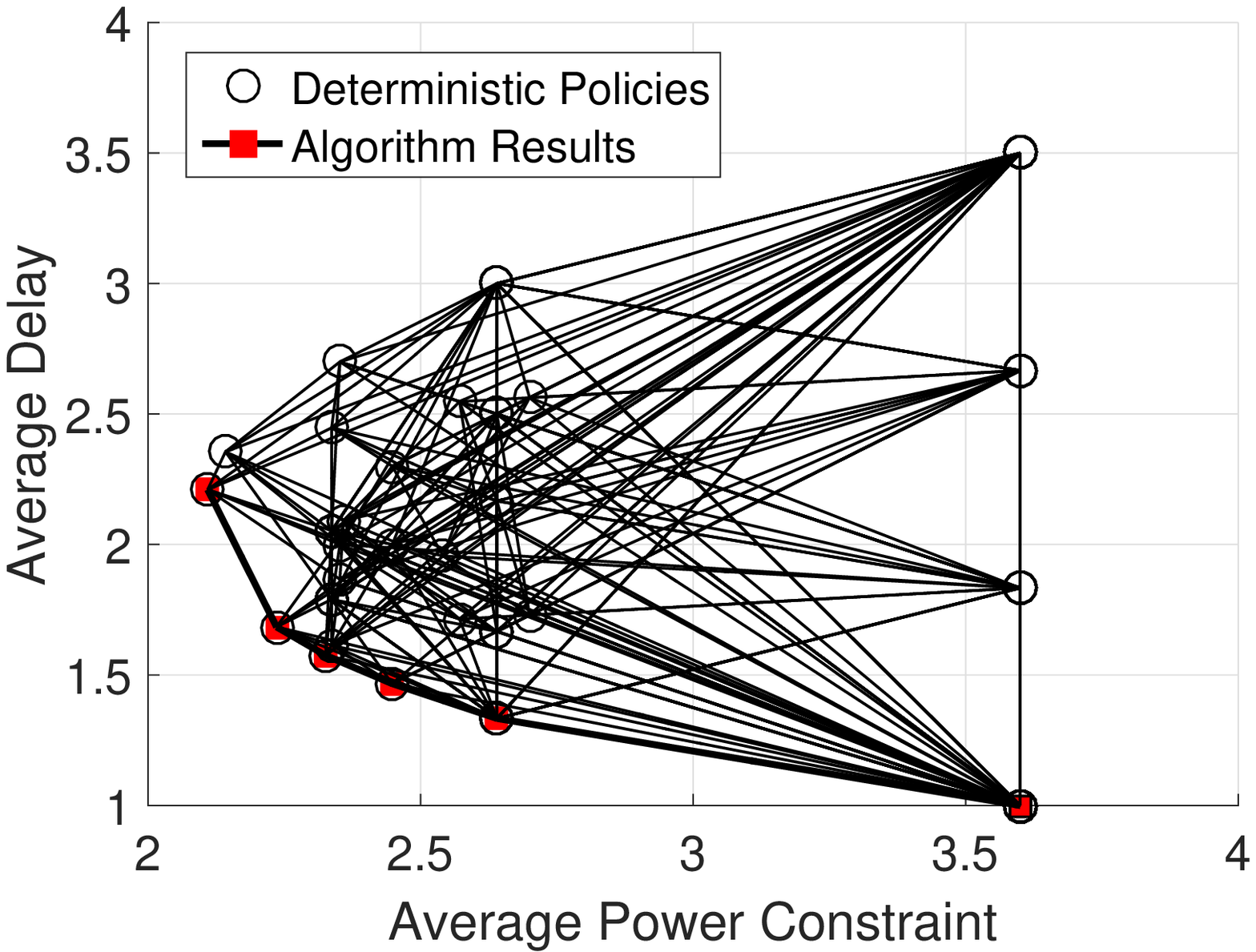}
\caption{Points Corresponding to Deterministic Policies \newline and Generated Polygons in the Delay-Power Plane}
\label{fig_policy}
\end{minipage}% 
\begin{minipage}[b]{0.5\textwidth} 
\centering
\includegraphics[width=1\columnwidth]{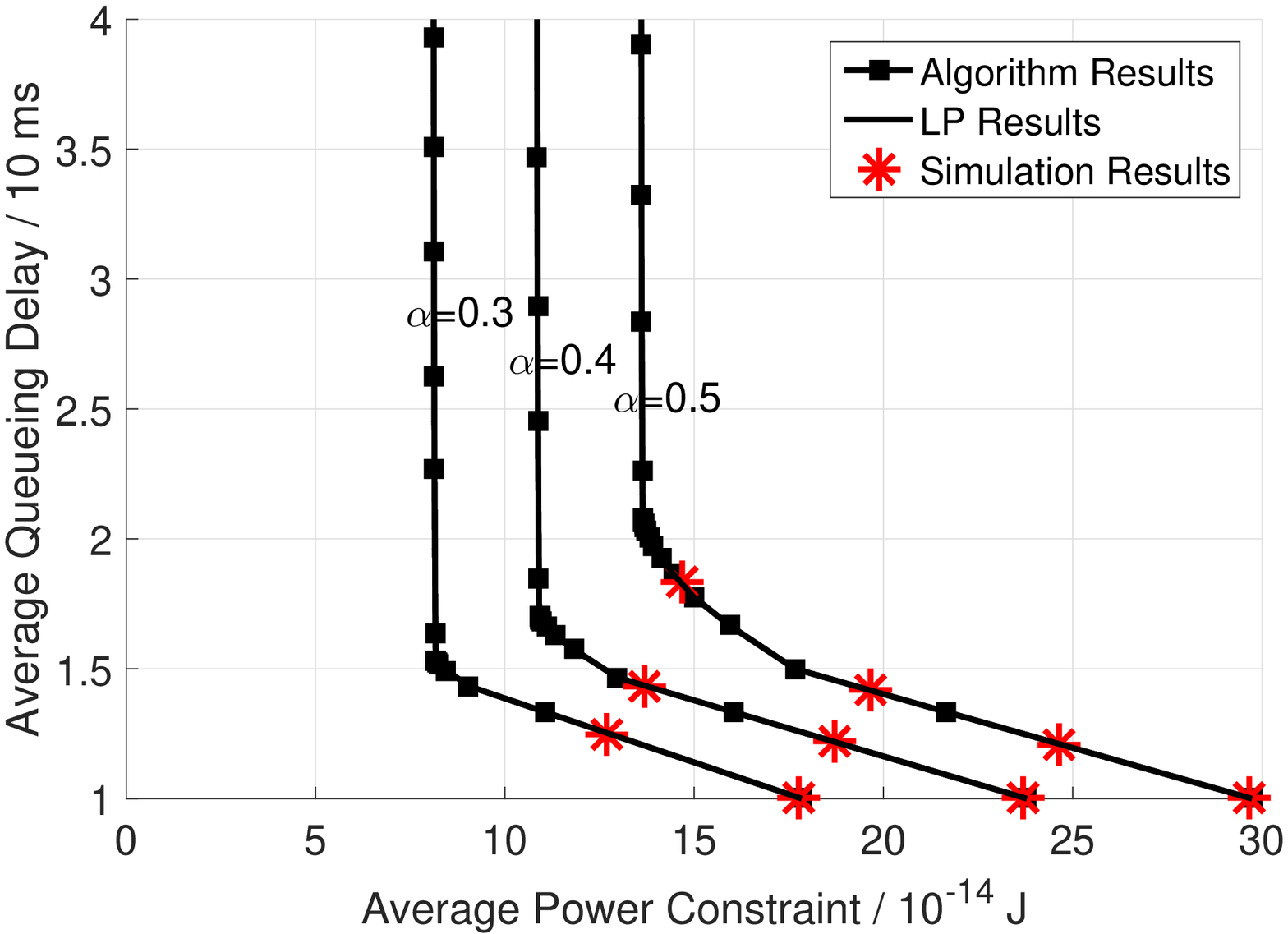}
\caption{Optimal Delay-Power Tradeoff Curves}
\label{fig_optimization}
\end{minipage}
\vspace{-0.8cm}
\end{figure}

In \figurename~\ref{fig_policy}, we plot all the delay-power points generated by deterministic policies, and connect the points whose corresponding policies take different actions in only one state. By conducting this operation, any generated polygon, a concept introduced in Appendix \ref{proof_region=polygon}, is contained in the figure. For any two deterministic policies, there is a convex polygon generated by them. Any point inside a generated polygon can be obtained by a policy. As we can see, the feasible delay-power region is made up of all the generated polygons. According to our proof in Appendix \ref{proof_region=polygon}, the feasible region is covered by basic polygons. The parameters for this figure are $Q=6$, $A=3$, $M=3$, $\alpha=0.4$, $P_0=0$, $P_1=1$, $P_2=4$, $P_3=9$. As proven in Theorem \ref{theorem_region=polygon}, the feasible delay-power region is the convex hull of all the points obtained by deterministic policies. The optimal delay-power tradeoff curve is obtained by Algorithm \ref{algo_curve}. Therefore the vertices of the curve are all corresponding to threshold-based deterministic policies, and neighbouring vertices are obtained by policies different in only one state. Policies to obtain the points between vertices of the optimal tradeoff curve are mixture of two deterministic policies.

The optimal delay-power tradeoff curves are demonstrated in \figurename~\ref{fig_optimization}, which are validated by Linear Programming and simulations. We consider a more practical scenario with adaptive M-PSK transmissions. The optional modulations are BPSK, QPSK, and 8-PSK. Assume the bandwidth = 1 MHz, the length of a timeslot = 10 ms, and the target bit error rate ber=$10^{-5}$. Set a data packet contains 10,000 bits. Then by adaptively applying BPSK, QPSK, or 8-PSK, we can respectively transmit 1, 2, or 3 packets in a timeslot, which means $S=3$. Assume the one-sided noise power spectral density $N_0$=-150 dBm/Hz. Then the transmission power for different transmission rates can be calculated as $P_0=0$ J, $P_1=9.0*10^{-14}$ J, $P_2=18.2*10^{-14}$ J, and $P_3=59.5*10^{-14}$ J. Assume in a timeslot, data arrive as a Bernoulli process. Each arrival contains $A=3$ packets. Set the buffer size $Q=100$. The optimal delay-power tradeoff curves are shown in \figurename~\ref{fig_optimization}, with $\alpha=0.3$, 0.4, and 0.5 respectively. It is demonstrated that the optimal delay-power tradeoff curves obtained by Linear Programming completely overlap the optimal tradeoff curves generated by Algorithm \ref{algo_curve}. The results are further validated by simulations, which are shown in ``*'' markers. As proven in Corollary \ref{corollary_piecewiselinear}, the optimal tradeoff curves are piecewise linear, decreasing, and convex. The vertices of the curves are marked by squares. The corresponding optimal policies can be checked as threshold-based. With $\alpha$ increasing, the curve gets higher because of the heavier workload. The minimum average delay is 1 for all curves, because when we transmit as much as we can, all data packets will stay in the queue for exactly one timeslot. The curve gets very steep when the power constraint decreases. This is because, when the power constraint gets tighter, we will mainly transmit with BPSK and QPSK. Since $P_1 \approx  \frac{P_2}{2}$, different policies will have similar average power consumption.

\section{Conclusion}
In this paper, we obtain the optimal tradeoff between the average delay and the average power consumption in a communication system. The transmission for each timeslot is scheduled according to the buffer state, considering an average power constraint. This problem is formulated as a Constrained Markov Decision Process. We first study the Lagrangian relaxation of the CMDP problem, and prove that it has deterministic threshold-based optimal policies. Then, we show that the feasible delay-power region is a convex polygon, and the optimal delay-power tradeoff curve is piecewise linear, whose vertices are obtained by the optimal solution to the relaxation problem, and the neighbouring vertices of which are obtained by policies taking different actions in only one state. Based on these results, the optimal policies for the CMDP are proven to be threshold-based, and we propose an efficient algorithm to obtain the optimal power-delay trade-off. The theoretical results and the proposed algorithm are validated by Linear Programming and simulations.

\appendices

\section{Proof of Theorem \ref{theorem_closed_class}}
Denote the set of transient states which have access to $\mathcal{C}_l$ as $\mathcal{C}_l^t$. Denote the set of transient states which don't have access to $\mathcal{C}_l$ as $\mathcal{C}_{nl}^t$. Therefore $\{\mathcal{C}_1,\cdots,\mathcal{C}_L,\mathcal{C}_l^t,\mathcal{C}_{nl}^t\}$ is a partition of all the states. It is straightforward that there should be at least one state $c\in\bigcup_{i=1,i\neq l}^{+\infty} \mathcal{C}_i \cup \mathcal{C}_{nl}^t$ which is adjacent to a state $c'\in \mathcal{C}_l \cup \mathcal{C}_l^t$, which means $|c-c'|=1$. If $c-c'=1$, we transmit 1 packet in state $c$; if $c'-c=1$, we transmit $(A-1)$ packet(s) in state $c$. Therefore we can always modify the transmission policy for state $c$ so that state $c$ can access $\mathcal{C}_l \cup \mathcal{C}_l^t$. Then $c$ will be a transient state which has access to $\mathcal{C}_l$, and so are the states which communicate with $c$.

Renew the partition since the state transition is changed. According to the above operation, the set $\mathcal{C}_l$ won't change, but $\mathcal{C}_l^t$ will be strictly increasing. Therefore, by repeating the same operation finite times, all the states will be partitioned in $\mathcal{C}_l$ and $\mathcal{C}_l^t$. Therefore $\mathcal{C}_l$ is its only closed communication class and the corresponding transmission policy is the $\boldsymbol{F}_l$ we request.

Since $\boldsymbol{F}$ and $\boldsymbol{F}_l$ still have the same policy for the states in $\mathcal{C}_l$, the limiting distribution and the average cost of the Markov chain generated by $\boldsymbol{F}$ starting from state $c\in\mathcal{C}_l$ are the same as the limiting distribution and the average cost of the Markov chain generated by $\boldsymbol{F}_l$.

\label{proof_closed_class}

\section{Proof of Lemma \ref{lemma_linearcombination}}
We will prove the two conclusions one by one.

1) From the definition of $\boldsymbol{H}_{\boldsymbol{F}}$ and $\boldsymbol{p}_{\boldsymbol{F}}$, we can see that if $\boldsymbol{F}''=(1-\epsilon)\boldsymbol{F}+\epsilon\boldsymbol{F}'$, then $\boldsymbol{H}_{\boldsymbol{F}''}=(1-\epsilon)\boldsymbol{H}_{\boldsymbol{F}}+\epsilon\boldsymbol{H}_{\boldsymbol{F}'}$ and $\boldsymbol{p}_{\boldsymbol{F}''}=(1-\epsilon)\boldsymbol{p}_{\boldsymbol{F}}+\epsilon\boldsymbol{p}_{\boldsymbol{F}'}$. Denote $\Delta\boldsymbol{H}=\boldsymbol{H}_{\boldsymbol{F}'}-\boldsymbol{H}_{\boldsymbol{F}}$ and $\Delta\boldsymbol{p}=\boldsymbol{p}_{\boldsymbol{F}'}-\boldsymbol{p}_{\boldsymbol{F}}$. Since $\boldsymbol{F}$ and $\boldsymbol{F}'$ are different only in the $(q+1)$th row, it can be derived that $\Delta \boldsymbol{H}$ has nonzero element only in the $(q+1)$th column, and the $(q+1)$th element of $\Delta \boldsymbol{p}$ is its only nonzero element. Therefore $\Delta \boldsymbol{H}$ can be denoted as $\left[\boldsymbol{0},\cdots,\boldsymbol{\delta}_q,\cdots,\boldsymbol{0}\right]$, where $\boldsymbol{\delta}_q$ is its $(q+1)$th column. $\Delta \boldsymbol{p}$ can be denoted as $\left[0,\cdots,\zeta_q,\cdots,0\right]^T$, where $\zeta_q$ is its $(q+1)$th element. Also, we denote $\boldsymbol{H}_{\boldsymbol{F}}^{-1}=\left[
\begin{array}{c}
\boldsymbol{h}_0^T\\
\boldsymbol{h}_1^T\\
\vdots\\
\boldsymbol{h}_Q^T
\end{array}
\right]$. Hence
$
(\boldsymbol{H}_{\boldsymbol{F}}^{-1}\Delta\boldsymbol{H})\boldsymbol{H}_{\boldsymbol{F}}^{-1}=\left[
\begin{array}{c}
(\boldsymbol{h}_0^T\boldsymbol{\delta}_q)\boldsymbol{h}_q^T\\
(\boldsymbol{h}_1^T\boldsymbol{\delta}_q)\boldsymbol{h}_q^T\\
\vdots\\
(\boldsymbol{h}_Q^T\boldsymbol{\delta}_q)\boldsymbol{h}_q^T
\end{array}
\right]
$.

By mathematical induction, we can have that for $i\ge 1$,
\begin{align}
(\boldsymbol{H}_{\boldsymbol{F}}^{-1}\Delta\boldsymbol{H})^{i}\boldsymbol{H}_{\boldsymbol{F}}^{-1}=\left[
\begin{array}{c}
(\boldsymbol{h}_0^T\boldsymbol{\delta}_q)(\boldsymbol{h}_q^T\boldsymbol{\delta}_q)^{i-1}\boldsymbol{h}_q^T\\
(\boldsymbol{h}_1^T\boldsymbol{\delta}_q)(\boldsymbol{h}_q^T\boldsymbol{\delta}_q)^{i-1}\boldsymbol{h}_q^T\\
\vdots\\
(\boldsymbol{h}_Q^T\boldsymbol{\delta}_q)(\boldsymbol{h}_q^T\boldsymbol{\delta}_q)^{i-1}\boldsymbol{h}_q^T
\end{array}
\right]=(\boldsymbol{h}_q^T\boldsymbol{\delta}_q)^{i-1}(\boldsymbol{H}_{\boldsymbol{F}}^{-1}\Delta\boldsymbol{H})\boldsymbol{H}_{\boldsymbol{F}}^{-1},
\end{align}
and
$\Delta\boldsymbol{p}^T\boldsymbol{H}_{\boldsymbol{F}}^{-1}(\boldsymbol{H}_{\boldsymbol{F}}^{-1}\Delta\boldsymbol{H})^{i-1}=\zeta_q(\boldsymbol{h}_q^T\boldsymbol{\delta}_q)^{i-1}\boldsymbol{h}_q^T$.

Therefore the expansion $(\boldsymbol{H}_{\boldsymbol{F}}+\epsilon\Delta\boldsymbol{H})^{-1}
%=&\sum_{i=0}^{+\infty} (-\epsilon)^{i}(\boldsymbol{H}_{\boldsymbol{F}}^{-1}\Delta\boldsymbol{H})^{i}\boldsymbol{H}_{\boldsymbol{F}}^{-1}\\
=\boldsymbol{H}_{\boldsymbol{F}}^{-1}
+\sum_{i=1}^{+\infty}(-\epsilon)^i(\boldsymbol{h}_q^T\boldsymbol{\delta}_q)^{i-1}(\boldsymbol{H}_{\boldsymbol{F}}^{-1}\Delta\boldsymbol{H})\boldsymbol{H}_{\boldsymbol{F}}^{-1}$.

From (\ref{pi-H}), (\ref{PwithPi}) and (\ref{DwithPi}), we have $P_{\boldsymbol{F}}=\boldsymbol{p}_{\boldsymbol{F}}^T\boldsymbol{H}_{\boldsymbol{F}}^{-1}\boldsymbol{c}$ and $D_{\boldsymbol{F}}=\frac{1}{\alpha A}\boldsymbol{d}^T\boldsymbol{H}_{\boldsymbol{F}}^{-1}\boldsymbol{c}-1$. Therefore
\begin{align}
&\frac{P_{\boldsymbol{F}''}-P_{\boldsymbol{F}}}{P_{\boldsymbol{F}'}-P_{\boldsymbol{F}}}=\frac{(\boldsymbol{p}_{\boldsymbol{F}}+\epsilon\Delta\boldsymbol{p})^T(\boldsymbol{H}_{\boldsymbol{F}}+\epsilon\Delta\boldsymbol{H})^{-1}\boldsymbol{c}-\boldsymbol{p}_{\boldsymbol{F}}^T\boldsymbol{H}_{\boldsymbol{F}}^{-1}\boldsymbol{c}}{(\boldsymbol{p}_{\boldsymbol{F}}+\Delta\boldsymbol{p})^T(\boldsymbol{H}_{\boldsymbol{F}}+\Delta\boldsymbol{H})^{-1}\boldsymbol{c}-\boldsymbol{p}_{\boldsymbol{F}}^T\boldsymbol{H}_{\boldsymbol{F}}^{-1}\boldsymbol{c}}\\
%=&\frac{
%\boldsymbol{p}_{\boldsymbol{F}}^T\left[(\boldsymbol{H}_{\boldsymbol{F}}+\epsilon\Delta\boldsymbol{H})^{-1}-\boldsymbol{H}_{\boldsymbol{F}}^{-1}\right]\boldsymbol{c}
%+\epsilon\Delta\boldsymbol{p}^T(\boldsymbol{H}_{\boldsymbol{F}}+\epsilon\Delta\boldsymbol{H})^{-1}\boldsymbol{c}
%}
%{
%\boldsymbol{p}_{\boldsymbol{F}}^T\left[(\boldsymbol{H}_{\boldsymbol{F}}+\Delta\boldsymbol{H})^{-1}-\boldsymbol{H}_{\boldsymbol{F}}^{-1}\right]\boldsymbol{c}
%+\Delta\boldsymbol{p}^T(\boldsymbol{H}_{\boldsymbol{F}}+\Delta\boldsymbol{H})^{-1}\boldsymbol{c}
%}\\
=&\frac{
\boldsymbol{p}_{\boldsymbol{F}}^T\left[\sum_{i=1}^{+\infty}(-\epsilon)^i(\boldsymbol{h}_q^T\boldsymbol{\delta}_q)^{i-1}(\boldsymbol{H}_{\boldsymbol{F}}^{-1}\Delta\boldsymbol{H})\boldsymbol{H}_{\boldsymbol{F}}^{-1}\right]\boldsymbol{c}
-\Delta\boldsymbol{p}^T\left[\sum_{i=1}^{+\infty} (-\epsilon)^{i}(\boldsymbol{H}_{\boldsymbol{F}}^{-1}\Delta\boldsymbol{H})^{i-1}\boldsymbol{H}_{\boldsymbol{F}}^{-1}\right]\boldsymbol{c}
}
{
\boldsymbol{p}_{\boldsymbol{F}}^T\left[\sum_{i=1}^{+\infty}(-1)^i(\boldsymbol{h}_q^T\boldsymbol{\delta}_q)^{i-1}(\boldsymbol{H}_{\boldsymbol{F}}^{-1}\Delta\boldsymbol{H})\boldsymbol{H}_{\boldsymbol{F}}^{-1}\right]\boldsymbol{c}
-\Delta\boldsymbol{p}^T\left[\sum_{i=1}^{+\infty} (-1)^{i}(\boldsymbol{H}_{\boldsymbol{F}}^{-1}\Delta\boldsymbol{H})^{i-1}\boldsymbol{H}_{\boldsymbol{F}}^{-1}\right]\boldsymbol{c}
}\\
%=&\frac{
%\sum_{i=1}^{+\infty}(-\epsilon)^i(\boldsymbol{h}_q^T\boldsymbol{\delta}_q)^{i-1}\boldsymbol{p}_{\boldsymbol{F}}^T(\boldsymbol{H}_{\boldsymbol{F}}^{-1}\Delta\boldsymbol{H})\boldsymbol{H}_{\boldsymbol{F}}^{-1}\boldsymbol{c}
%-\sum_{i=1}^{+\infty} (-\epsilon)^{i}\zeta_q(\boldsymbol{h}_q^T\boldsymbol{\delta}_q)^{i-1}\boldsymbol{h}_q^T\boldsymbol{c}
%}
%{
%\sum_{i=1}^{+\infty}(-1)^i(\boldsymbol{h}_q^T\boldsymbol{\delta}_q)^{i-1}\boldsymbol{p}_{\boldsymbol{F}}^T(\boldsymbol{H}_{\boldsymbol{F}}^{-1}\Delta\boldsymbol{H})\boldsymbol{H}_{\boldsymbol{F}}^{-1}\boldsymbol{c}
%-\sum_{i=1}^{+\infty} (-1)^{i}\zeta_q(\boldsymbol{h}_q^T\boldsymbol{\delta}_q)^{i-1}\boldsymbol{h}_q^T\boldsymbol{c}
%}\\
=&\frac{\sum_{i=1}^{+\infty}(-\epsilon)^i(\boldsymbol{h}_q^T\boldsymbol{\delta}_q)^{i-1}}
{\sum_{i=1}^{+\infty}(-1)^i(\boldsymbol{h}_q^T\boldsymbol{\delta}_q)^{i-1}}
=\frac{\epsilon+\epsilon\boldsymbol{h}_q^T\boldsymbol{\delta}_q}{1+\epsilon\boldsymbol{h}_q^T\boldsymbol{\delta}_q}
\end{align}
and
\begin{align}
& \frac{D_{\boldsymbol{F}''}-D_{\boldsymbol{F}}}{D_{\boldsymbol{F}'}-D_{\boldsymbol{F}}}=\frac{\boldsymbol{d}^T(\boldsymbol{H}_{\boldsymbol{F}}+\epsilon\Delta\boldsymbol{H})^{-1}\boldsymbol{c}-\boldsymbol{d}^T\boldsymbol{H}_{\boldsymbol{F}}^{-1}\boldsymbol{c}}{\boldsymbol{d}^T(\boldsymbol{H}_{\boldsymbol{F}}+\Delta\boldsymbol{H})^{-1}\boldsymbol{c}-\boldsymbol{d}^T\boldsymbol{H}_{\boldsymbol{F}}^{-1}\boldsymbol{c}}\\
=&\frac{\boldsymbol{d}^T(\sum_{i=1}^{+\infty}(-\epsilon)^i(\boldsymbol{h}_q^T\boldsymbol{\delta}_q)^{i-1}(\boldsymbol{H}_{\boldsymbol{F}}^{-1}\Delta\boldsymbol{H})\boldsymbol{H}_{\boldsymbol{F}}^{-1})\boldsymbol{c}}{\boldsymbol{d}^T(\sum_{i=1}^{+\infty}(-1)^i(\boldsymbol{h}_q^T\boldsymbol{\delta}_q)^{i-1}(\boldsymbol{H}_{\boldsymbol{F}}^{-1}\Delta\boldsymbol{H})\boldsymbol{H}_{\boldsymbol{F}}^{-1})\boldsymbol{c}}
=\frac{\sum_{i=1}^{+\infty}(-\epsilon)^i(\boldsymbol{h}_q^T\boldsymbol{\delta}_q)^{i-1}}{\sum_{i=1}^{+\infty}(-1)^i(\boldsymbol{h}_q^T\boldsymbol{\delta}_q)^{i-1}}
=\frac{\epsilon+\epsilon\boldsymbol{h}_q^T\boldsymbol{\delta}_q}{1+\epsilon\boldsymbol{h}_q^T\boldsymbol{\delta}_q}.
\end{align}
Hence $\frac{P_{\boldsymbol{F}''}-P_{\boldsymbol{F}}}{P_{\boldsymbol{F}'}-P_{\boldsymbol{F}}}=\frac{D_{\boldsymbol{F}''}-D_{\boldsymbol{F}}}{D_{\boldsymbol{F}'}-D_{\boldsymbol{F}}}=\frac{\epsilon+\epsilon\boldsymbol{h}_q^T\boldsymbol{\delta}_q}{1+\epsilon\boldsymbol{h}_q^T\boldsymbol{\delta}_q}=\epsilon'$, so that $P_{\boldsymbol{F}''}=(1-\epsilon')P_{\boldsymbol{F}}+\epsilon' P_{\boldsymbol{F}'}$ and $D_{\boldsymbol{F}''}=(1-\epsilon')D_{\boldsymbol{F}}+\epsilon' D_{\boldsymbol{F}'}$. Moreover, it can be observed that $\epsilon'=\frac{\epsilon+\epsilon\boldsymbol{h}_q^T\boldsymbol{\delta}_q}{1+\epsilon\boldsymbol{h}_q^T\boldsymbol{\delta}_q}$ is a continuous nondecreasing function.

2) From the first part, we know $\frac{P_{\boldsymbol{F}''}-P_{\boldsymbol{F}}}{P_{\boldsymbol{F}'}-P_{\boldsymbol{F}}}=\frac{D_{\boldsymbol{F}''}-D_{\boldsymbol{F}}}{D_{\boldsymbol{F}'}-D_{\boldsymbol{F}}}=\epsilon'$ and $\epsilon'$ is a continuous nondecreasing function of $\epsilon$. When $\epsilon=0$, we have $\epsilon'=0$. When $\epsilon=1$, we have $\epsilon'=1$. Therefore when $\epsilon$ changes from 0 to 1, the point $(P_{\boldsymbol{F}''},D_{\boldsymbol{F}''})$ moves on the line segment from $(P_{\boldsymbol{F}},D_{\boldsymbol{F}})$ to $(P_{\boldsymbol{F}'},D_{\boldsymbol{F}'})$. The slope of the line is
\begin{align}
&\frac{D_{\boldsymbol{F}'}-D_{\boldsymbol{F}}}{P_{\boldsymbol{F}'}-P_{\boldsymbol{F}}}=\frac{
\frac{1}{\alpha A}\boldsymbol{d}^T(\boldsymbol{H}_{\boldsymbol{F}}+\Delta\boldsymbol{H})^{-1}\boldsymbol{c}-\frac{1}{\alpha A}\boldsymbol{d}^T\boldsymbol{H}_{\boldsymbol{F}}^{-1}\boldsymbol{c}
}{
(\boldsymbol{p}_{\boldsymbol{F}}+\Delta\boldsymbol{p})^T(\boldsymbol{H}_{\boldsymbol{F}}+\Delta\boldsymbol{H})^{-1}\boldsymbol{c}-\boldsymbol{p}_{\boldsymbol{F}}^T\boldsymbol{H}_{\boldsymbol{F}}^{-1}\boldsymbol{c}
}\\
%=&\frac{\frac{1}{\alpha A}\boldsymbol{d}^T(\sum_{i=1}^{+\infty}(-1)^i(\boldsymbol{h}_q^T\boldsymbol{\delta}_q)^{i-1}(\boldsymbol{H}_{\boldsymbol{F}}^{-1}\Delta\boldsymbol{H})\boldsymbol{H}_{\boldsymbol{F}}^{-1})\boldsymbol{c}}
%{
%\sum_{i=1}^{+\infty}(-1)^i(\boldsymbol{h}_q^T\boldsymbol{\delta}_q)^{i-1}\boldsymbol{p}_{\boldsymbol{F}}^T(\boldsymbol{H}_{\boldsymbol{F}}^{-1}\Delta\boldsymbol{H})\boldsymbol{H}_{\boldsymbol{F}}^{-1}\boldsymbol{c}
%-\sum_{i=1}^{+\infty} (-1)^{i}\zeta_q(\boldsymbol{h}_q^T\boldsymbol{\delta}_q)^{i-1}\boldsymbol{h}_q^T\boldsymbol{c}
%}\\
=&\frac{\frac{1}{\alpha A}\boldsymbol{d}^T\boldsymbol{H}_{\boldsymbol{F}}^{-1}\Delta\boldsymbol{H}\boldsymbol{H}_{\boldsymbol{F}}^{-1}\boldsymbol{c}}
{\boldsymbol{p}_{\boldsymbol{F}}^T\boldsymbol{H}_{\boldsymbol{F}}^{-1}\Delta\boldsymbol{H}\boldsymbol{H}_{\boldsymbol{F}}^{-1}\boldsymbol{c}-\zeta_q\boldsymbol{h}_q^T\boldsymbol{c}}=\frac{\boldsymbol{d}^T\boldsymbol{H}_{\boldsymbol{F}}^{-1}\boldsymbol{\delta}_q}
{\alpha A (\boldsymbol{p}_{\boldsymbol{F}}^T\boldsymbol{H}_{\boldsymbol{F}}^{-1}\boldsymbol{\delta}_q-\zeta_q)}.\label{slope}
\end{align}

\label{proof_linearcombination}

\section{Proof of Theorem \ref{theorem_region=polygon}}

\begin{figure*}[t]
\centering
\subfloat[A Convex Basic Polygon in the Normal Shape]{\includegraphics[width=0.45\columnwidth]{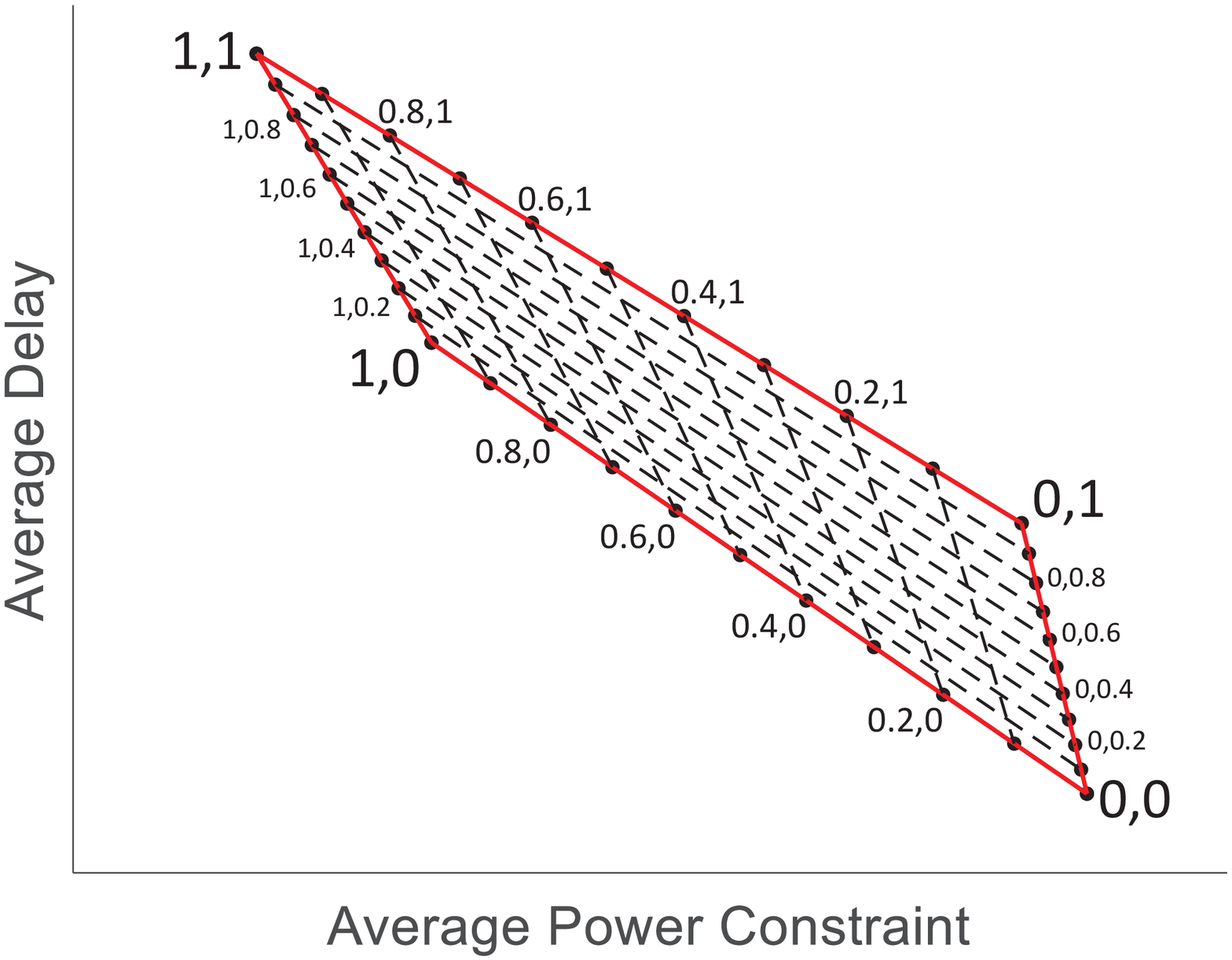}
\label{fig_basic_1}}
\subfloat[A Nonconvex Basic Polygon in the Boomerang Shape]{\includegraphics[width=0.45\columnwidth]{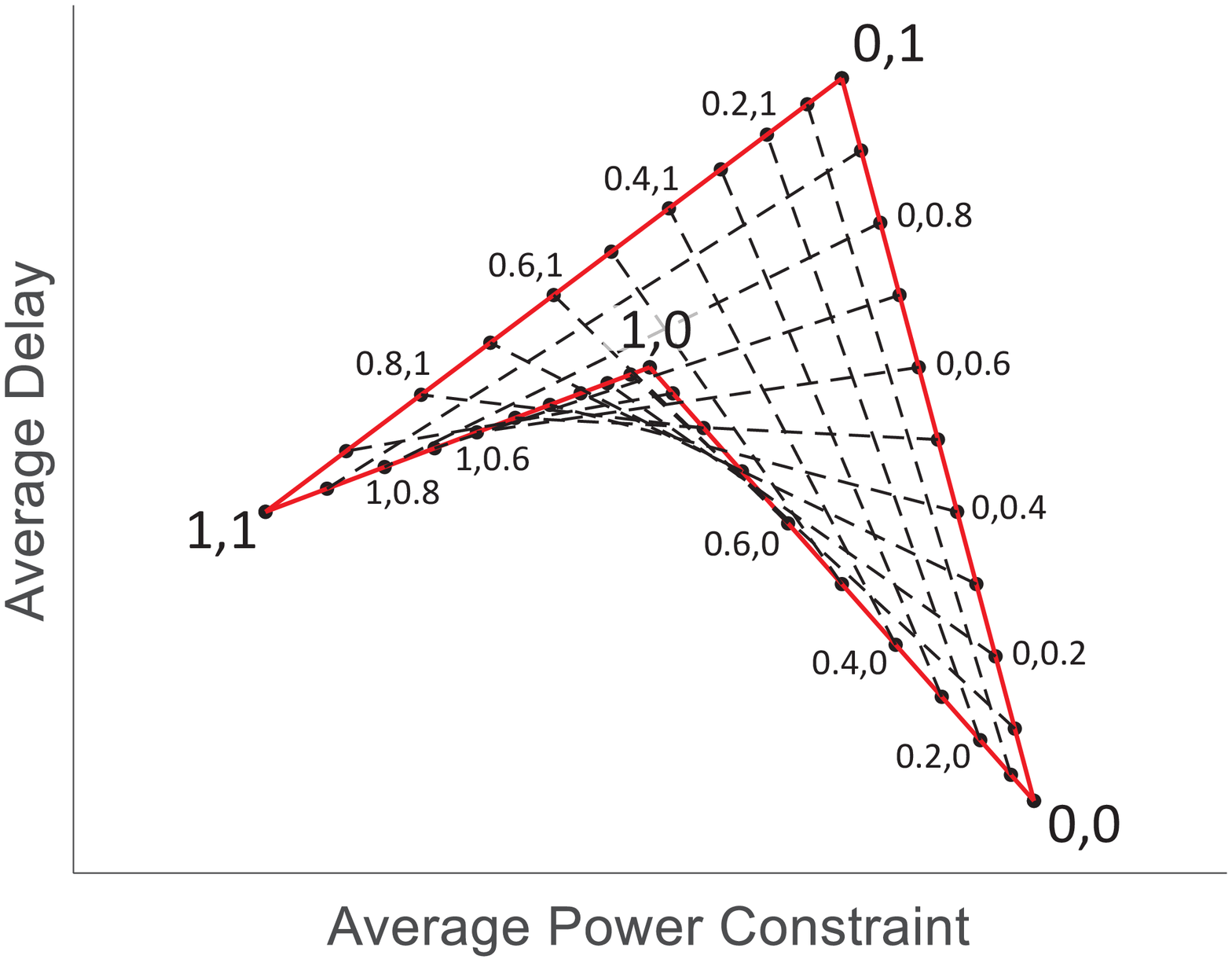}
\label{fig_basic_2}}\\
\subfloat[A NonConvex Basic Polygon in the Butterfly Shape]{\includegraphics[width=0.45\columnwidth]{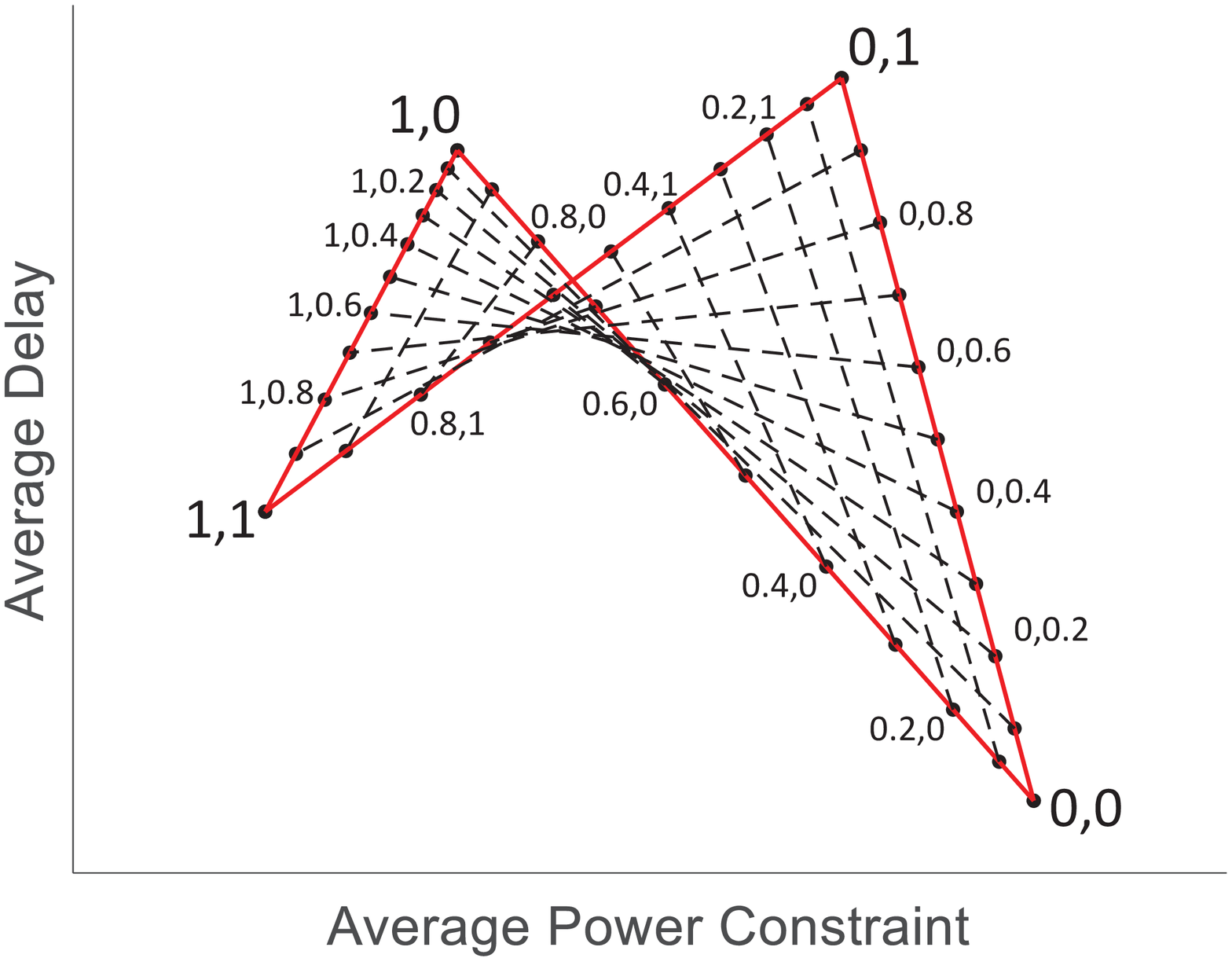}
\label{fig_basic_3}}
\subfloat[A Nonconvex Basic Polygon in the Slender Butterfly Shape]{\includegraphics[width=0.45\columnwidth]{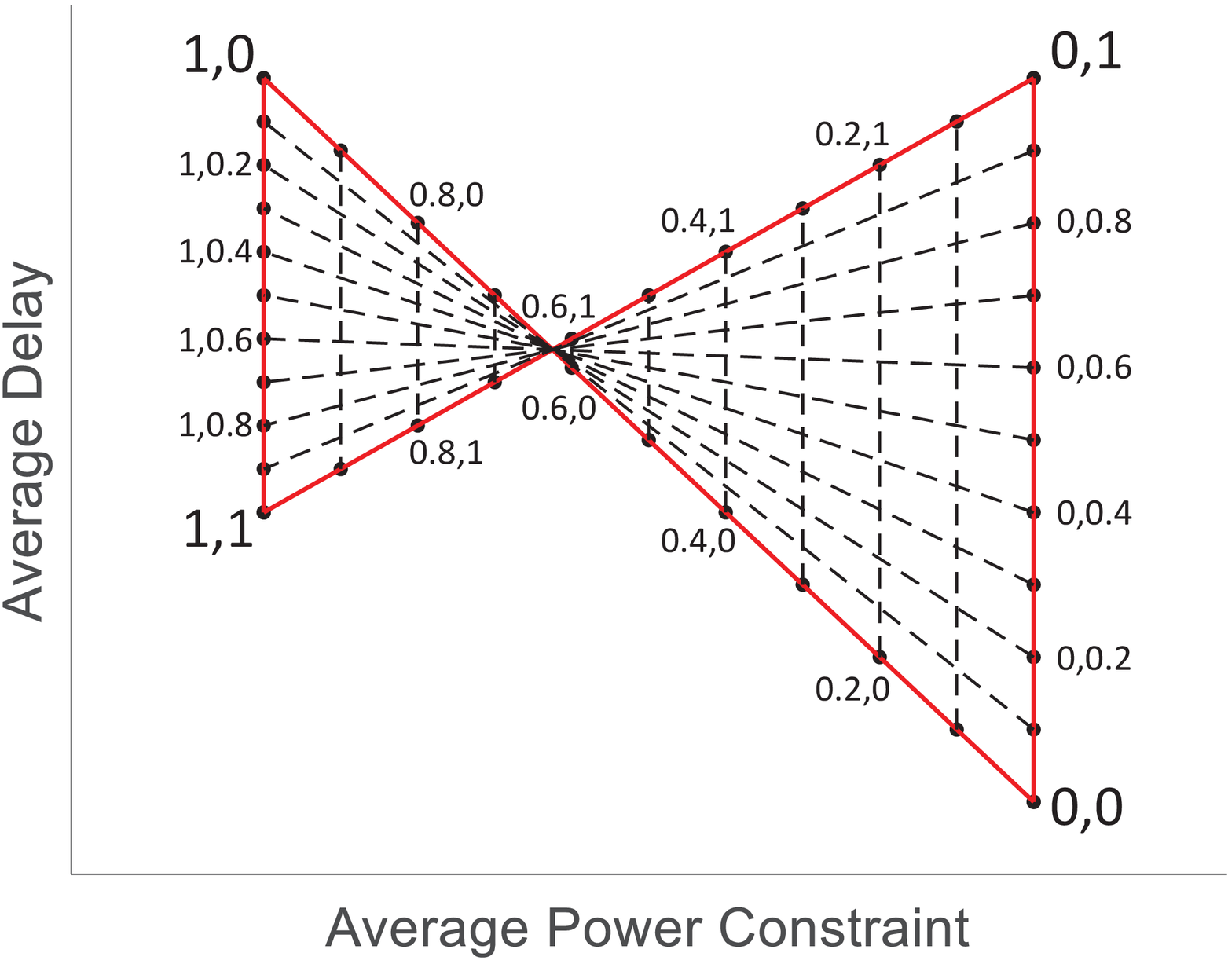}
\label{fig_basic_4}}
\caption{Demonstration for Basic Polygons}
\label{fig_basic}
\vspace{-0.8cm}
\end{figure*}

Denote $\mathcal{C}=\textbf{conv } \{Z_{\boldsymbol{F}} | \boldsymbol{F} \in \mathcal{F}_D \}$ as the convex hull of points in the delay-power plane corresponding to deterministic scheduling policies. By proving $\mathcal{R}=\mathcal{C}$, we can have that $\mathcal{R}$ is a convex polygon whose vertices are all obtained by deterministic scheduling policies.

The proof contains three parts. In the first part, we will prove $\mathcal{R}\subseteq\mathcal{C}$ by the construction method. In the second part, we define the concepts of basic polygons and compound polygons, and prove that they are convex, based on which $\mathcal{R}\supseteq\mathcal{C}$ can be proven. By combining the results in these two parts, we will have $\mathcal{R}=\mathcal{C}$. Finally, in the third part, we will prove the policies corresponding to neighbouring vertices of $\mathcal{R}$ are different in only one state.

\textbf{Part I. Prove $\mathcal{R}\subseteq\mathcal{C}$}

For any specific probabilistic policy $\boldsymbol{F}$ where $0<f_{q^*,s^*}<1$, we construct

$\boldsymbol{F}'=
\begin{cases}
f'_{q,s}=1 & q=q^*,s=s^*\\
f'_{q,s}=0 & q=q^*,s \neq s^*\\
f'_{q,s}=f_{q,s} & \text{else}
\end{cases}$ \quad and \quad $
\boldsymbol{F}''=
\begin{cases}
f''_{q,s}=0 & q=q^*,s=s^*\\
f''_{q,s}=\frac{f_{q,s}}{1-f_{q^*,s^*}} & q=q^*,s \neq s^*\\
f''_{q,s}=f_{q,s} & \text{else}
\end{cases}$.

Since $0 \le \frac{f_{q,s}}{1-f_{q^*,s^*}} \le 1$, and whenever $f_{q,s}=0$, we have $f'_{q,s}=f''_{q,s}=0$, the constructed policies $\boldsymbol{F}'$ and $\boldsymbol{F}''$ are feasible. It can be seen that $\boldsymbol{F}=f_{q^*,s^*}\boldsymbol{F}'+(1-f_{q^*,s^*})\boldsymbol{F}''$. Since $\boldsymbol{F}$ is a convex combination of $\boldsymbol{F}'$ and $\boldsymbol{F}''$, also $\boldsymbol{F}'$ and $\boldsymbol{F}''$ are only different in the $(q^*+1)$th row, from Lemma \ref{lemma_linearcombination}, we know that $Z_{\boldsymbol{F}}$ is a convex combination of $Z_{\boldsymbol{F}'}$ and $Z_{\boldsymbol{F}''}$. Note that $f'_{q^*,s^*}$ and $f''_{q^*,s^*}$ are integers. Also, in $\boldsymbol{F}'$ and $\boldsymbol{F}''$, no new decimal elements will be introduced. Hence we can conclude that, in finite steps, the point $Z_{\boldsymbol{F}}$ can be expressed as a convex combination of points in the delay-power plane corresponding to deterministic scheduling policies, which means $Z_{\boldsymbol{F}}\in\mathcal{C}$. From the arbitrariness of $\boldsymbol{F}$, we can see $\mathcal{R}\subseteq\mathcal{C}$ is proven.

\textbf{Part II. Prove $\mathcal{R}\supseteq\mathcal{C}$}

In the second part, we will first define the concepts of basic polygons and compound polygons in Part II.0. Then basic polygons and compound polygons will be proven convex in Part II.1 and Part II.2 respectively. Based on these results, we will prove $\mathcal{R}\supseteq\mathcal{C}$ in Part II.3.

\textbf{Part II.0 Introduce the Concepts of Basic Polygons and Compound Polygons}

\begin{figure}[t]
\centering
\subfloat[A Convex Compound Polygon]{\includegraphics[width=0.3\columnwidth]{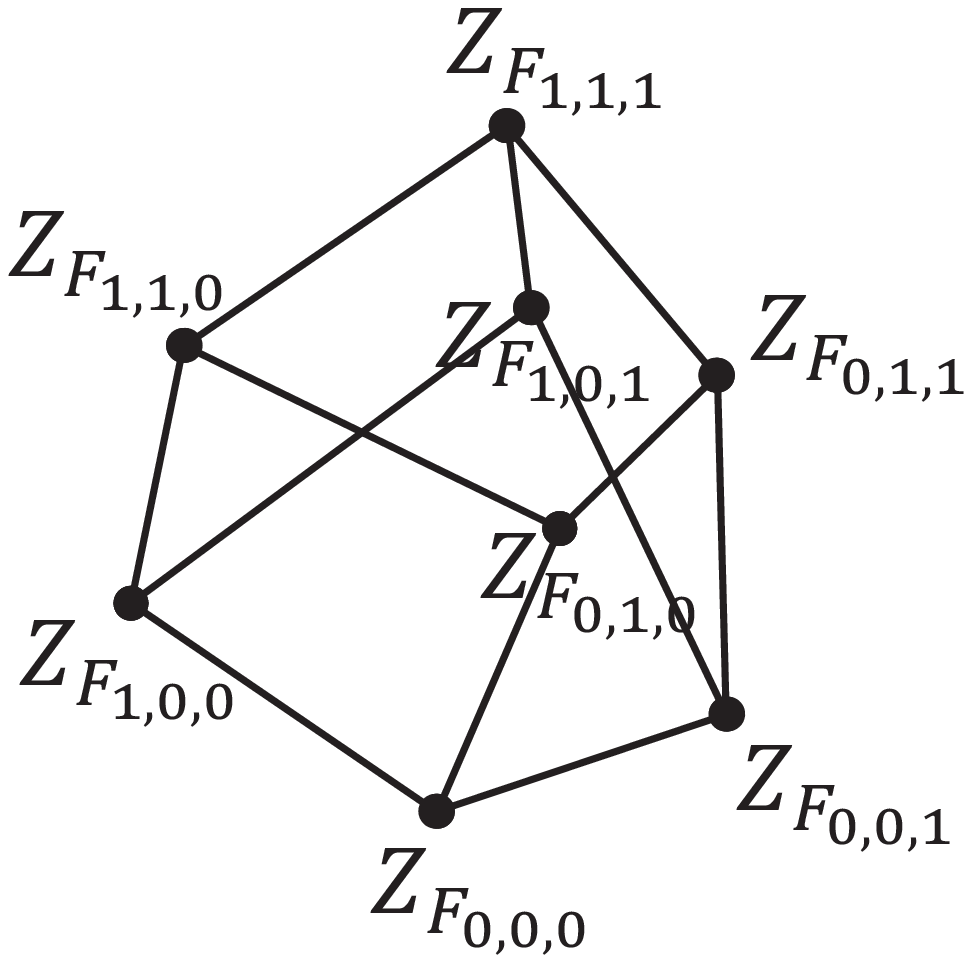}
\label{fig_convex_1}}
\qquad\qquad
\subfloat[A Nonconvex Compound Polygon]{\includegraphics[width=0.2\columnwidth]{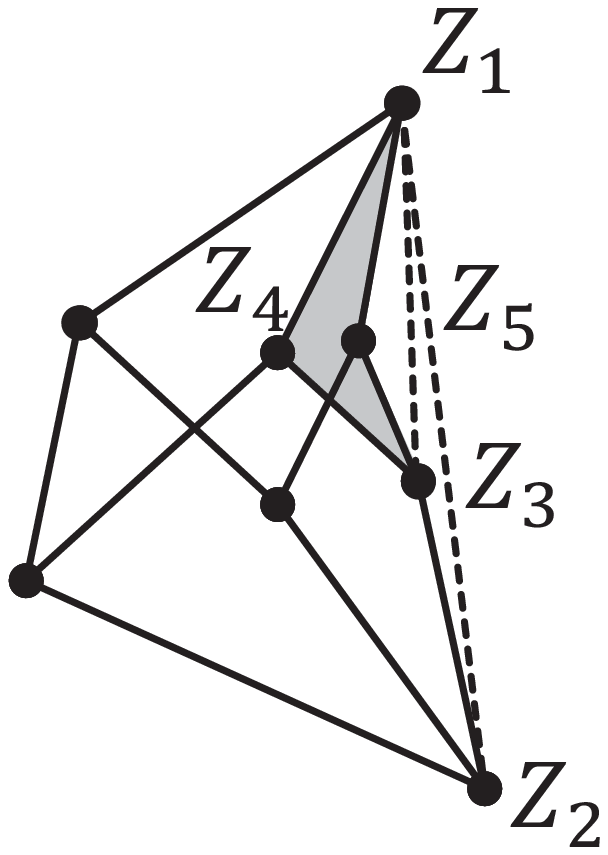}
\label{fig_convex_2}}
\caption{Demonstration for Compound Polygons}
\label{fig_convex}
\vspace{-0.8cm}
\end{figure}

For two deterministic policies $\boldsymbol{F}$ and $\boldsymbol{F}'$ which are different in $K$ states, namely $q_1,\cdots,q_K$, define
$\boldsymbol{F}_{b_1,b_2,\cdots,b_K}(q,:)=\begin{cases}
(1-b_k)\boldsymbol{F}(q,:)+b_k\boldsymbol{F}'(q,:) & q=q_k,\\
\boldsymbol{F}(q,:) & q\neq q_1,\cdots,q_K,
\end{cases}$
where $0 \le b_k \le 1$ for all $k$. Thus $\boldsymbol{F}_{0,0,\cdots,0}=\boldsymbol{F}$, and $\boldsymbol{F}_{1,1,\cdots,1}=\boldsymbol{F}'$. With more $b_k$ equal to 0, the policy is more like $\boldsymbol{F}$. With more $b_k$ equal to 1, the policy is more like $\boldsymbol{F}$'. For policies $\boldsymbol{F}_{b_1,\cdots,b_k,\cdots,b_K}$ and $\boldsymbol{F}_{b_1,\cdots,b_k',\cdots,b_K}$ where $b_k \neq b_k'$, since they are different in only one state, according to Lemma \ref{lemma_linearcombination}, the delay-power point corresponding to their convex combination $Z_{\epsilon \boldsymbol{F}_{b_1,\cdots,b_k,\cdots,b_K} + (1-\epsilon)\boldsymbol{F}_{b_1,\cdots,b_k',\cdots,b_K}}$ is the convex combination of $Z_{\boldsymbol{F}_{b_1,\cdots,b_k,\cdots,b_K}}$ and $Z_{\boldsymbol{F}_{b_1,\cdots,b_k',\cdots,b_K}}$. However, for policies different in more than one state, the delay-power point corresponding to their convex combination is not necessarily the convex combination of their own delay-power points. Therefore, we introduce the concept of generated polygon to demonstrate the delay-power region of convex combinations of two policies. We plot $Z_{\boldsymbol{F}_{b_1,\cdots,b_K}}$, where $b_k=0$ or $1$ for all $1\le k \le K$, and connect the points whose corresponding policies are different in only one state. Therefore any point on any line segment can be obtained by a certain policy. We define the figure as a polygon generated by $\boldsymbol{F}$ and $\boldsymbol{F}'$. The red polygon in \figurename~\ref{fig_basic_1} and the polygon in \figurename~\ref{fig_convex_1} are demonstrations where $\boldsymbol{F}$ and $\boldsymbol{F}'$ are different in 2 and 3 states respectfully. If $K=2$, we call the polygon a basic polygon. If $K>2$, we call it a compound polygon. As demonstrated in \figurename~\ref{fig_convex_1}, a compound polygon contains multiple basic polygons.

\textbf{Part II.1 Prove a Basic Polygon is Convex and Any Point Inside a Basic Polygon can be Obtained by a Policy}

For better visuality, in \figurename~\ref{fig_basic}, we simplify the notation $Z_{\boldsymbol{F}_{b_1,b_2}}$ as $b_1,b_2$. According to different relative positions of $Z_{\boldsymbol{F}_{0,0}}$, $Z_{\boldsymbol{F}_{0,1}}$, $Z_{\boldsymbol{F}_{1,0}}$, and $Z_{\boldsymbol{F}_{1,1}}$, there are in total 3 possible shapes of basic polygons, as shown in \figurename~\ref{fig_basic_1}-\ref{fig_basic_3} respectfully. We name them as the normal shape, the boomerang shape, and the butterfly shape. The degenerate polygons such as triangles, line segments and points are considered included in the above three cases. Besides $\boldsymbol{F}_{b_1,b_2}$ with integral $b_1,b_2$ and the line segments connecting them, we also plot the points corresponding to policy $\boldsymbol{F}_{b_1,b_2}$ where one of $b_1,b_2$ is integer and the other one is decimal. We connect the points corresponding to policies with the same $b_1$ or $b_2$ in dashed lines. As demonstrated in \figurename~\ref{fig_basic}, we draw line segments $\overline{Z_{\boldsymbol{F}_{b_1,0}}Z_{\boldsymbol{F}_{b_1,1}}}$ where $b_1=0.1,0.2,\cdots,0.9$ and $\overline{Z_{\boldsymbol{F}_{0,b_2}}Z_{\boldsymbol{F}_{1,b_2}}}$ where $b_2=0.1,0.2,\cdots,0.9$. For any specific $b_1$ and $b_2$, the point $Z_{\boldsymbol{F}_{b_1,b_2}}$ should be on both $\overline{Z_{\boldsymbol{F}_{b_1,0}}Z_{\boldsymbol{F}_{b_1,1}}}$ and $\overline{Z_{\boldsymbol{F}_{0,b_2}}Z_{\boldsymbol{F}_{1,b_2}}}$. Because of the existence of $Z_{\boldsymbol{F}_{b_1,b_2}}$, line segments $\overline{Z_{\boldsymbol{F}_{b_1,0}}Z_{\boldsymbol{F}_{b_1,1}}}$ and $\overline{Z_{\boldsymbol{F}_{0,b_2}}Z_{\boldsymbol{F}_{1,b_2}}}$ should always have an intersection point for any specific $b_1$ and $b_2$. However, if there exist line segments outside the polygon, there exist $b_1$ and $b_2$ whose line segments don't intersect. Therefore, in the boomerang shape, there will always exist $b_1$ and $b_2$ whose line segments don't intersect. In the butterfly shape, there will exist $b_1$ and $b_2$ whose line segments don't intersect except the case that all the line segments are inside the basic polygon, as shown in \figurename~\ref{fig_basic_4}, which is named as the slender butterfly shape. In the slender butterfly shape, there exists a specific $b_1^*$ such that $\overline{Z_{\boldsymbol{F}_{b_1^*,0}}Z_{\boldsymbol{F}_{b_1^*,1}}}$ degenerates into a point, or there exists a specific $b_2^*$ such that $\overline{Z_{\boldsymbol{F}_{0,b_2^*}}Z_{\boldsymbol{F}_{1,b_2^*}}}$ degenerates into a point. Without loss of generality, we assume it is the $b_1^*$ case. It means that under policy $\boldsymbol{F}_{b_1^*,b_2}$, state $q_2$, the state corresponding to $b_2$, is a transient state. For $b_1\in(b_1^*-\epsilon,b_1^*+\epsilon)$ when $\epsilon$ is small enough, the Markov chain applying policy $F_{b_1,b_2}$ also has $q_2$ as a transient state, therefore $\overline{Z_{\boldsymbol{F}_{b_1,0}}Z_{\boldsymbol{F}_{b_1,1}}}$ also degenerates into a point. Thus $\overline{Z_{\boldsymbol{F}_{0,0}}Z_{\boldsymbol{F}_{1,0}}}$ and $\overline{Z_{\boldsymbol{F}_{0,1}}Z_{\boldsymbol{F}_{1,1}}}$ overlap, which means the slender butterfly shape always degenerates to a line segment, which can also be considered as a normal shape. Since the normal shape is the only possible shape of a basic polygon, the basic polygon is convex. Since the transition from the point $Z_{\boldsymbol{F}_{0,0}}$ to $Z_{\boldsymbol{F}_{1,1}}$ is termwise monotone and continuous, every point inside the basic polygon can be obtained by a policy.

\textbf{Part II.2 Prove a Compound Polygon is Convex}

For any two deterministic policies $\boldsymbol{F}$ and $\boldsymbol{F}'$, if their generated compound polygon is not convex, then there exist two vertices whose connecting line is outside the compound polygon, as demonstrated by $\overline{Z_1 Z_2}$ in \figurename~\ref{fig_convex_2}. Thus, there must exist two vertices who are connecting to the same point such that their connecting line is outside the compound polygon, as demonstrated by $\overline{Z_1 Z_3}$. The policy corresponding to these two vertices must be different in only two states, therefore there must be a basic polygon generated by them, as demonstrated by the filled polygon. Since $\overline{Z_1 Z_3}$ is outside the compound polygon, it is outside the basic polygon too, which is not possible because basic polygons are always convex. Therefore all generated compound polygons are convex.

\textbf{Part II.3 Prove $\mathcal{R}\supseteq\mathcal{C}$}

For any point $C\in\mathcal{C}$, it will surely fall into one of the compound polygons. Because otherwise, there will be at least one point corresponding to a deterministic policy which is outside any compound polygons, which is impossible. Any compound polygon is covered by basic polygons, therefore $C$ is inside at least one basic polygon. Since any point inside a basic polygon can be obtained by a policy, the point $C\in\mathcal{R}$. From the arbitrariness of $C$, we have $\mathcal{R}\supseteq\mathcal{C}$.

From Part II.1 and Part II.2, it can be seen that $\mathcal{R}=\mathcal{C}$. Since there are only finite deterministic policies in total, the set $\mathcal{R}$ is a convex polygon whose vertices are all obtained by deterministic scheduling policies.

\textbf{Part III. Neighbouring Vertices of $\mathcal{R}$}

For any two neighbouring vertices $Z_{\boldsymbol{F}}$ and $Z_{\boldsymbol{F}'}$ of $\mathcal{R}$, if $\boldsymbol{F}$ and $\boldsymbol{F}'$ are different in more than one state, their generated polygon is convex. If the line segment $\overline{Z_{\boldsymbol{F}}Z_{\boldsymbol{F}'}}$ is inside the generated polygon, $Z_{\boldsymbol{F}}$ and $Z_{\boldsymbol{F}'}$ are not neighbouring vertices. If the line segment $\overline{Z_{\boldsymbol{F}}Z_{\boldsymbol{F}'}}$ is on the boundary of the generated polygon, there will be other vertices between them, such that $Z_{\boldsymbol{F}}$ and $Z_{\boldsymbol{F}'}$ are not neighbouring, neither. Therefore, policies $\boldsymbol{F}$ and $\boldsymbol{F}'$ are deterministic and different in only one state.

\label{proof_region=polygon}

\section{Proof of Corollary \ref{corollary_piecewiselinear}}
\textbf{Monotonicity:}

Since $\mathcal{L}=\{(P,D)\in\mathcal{R}|\forall(P',D')\in\mathcal{R},\text{ either }P'\ge P\text{ or }D'\ge D\}$, for any $(P_1,D_1),(P_2,D_2)\in\mathcal{L}$ where $P_1<P_2$, we should have $D_1\ge D_2$. Therefore $\mathcal{L}$ is decreasing.

\textbf{Convexity:}

Since $\mathcal{R}$ is a convex polygon, for any $(P_1,D_1),(P_2,D_2)\in\mathcal{L}$, their convex combination is $(\theta P_1+(1-\theta) P_2,\theta D_1+(1-\theta) D_2)\in\mathcal{R}$. Therefore, there should be a point $(P_\theta,D_\theta)$ on $\mathcal{L}$ where $P_\theta = \theta P_1+(1-\theta) P_2$, and $D_\theta\le \theta D_1+(1-\theta) D_2$. Therefore $\mathcal{L}$ is convex.

\textbf{Piecewise Linearity:}

Since $\mathcal{R}$ is a convex polygon, it can be expressed as the intersection of a finite number of halfspaces, i.e., $\mathcal{R}=\bigcap_{i=1}^{I}\{(P,D)|a_i P+b_i D \ge c_i\}$. We divide $(a_i,b_i,c_i)$ into 2 categories according to the value of $a_i$ and $b_i$ as $(a_i^+,b_i^+,c_i^+)$ for $i=1,\cdots,I^+$ if $a_i>0$ and $b_i>0$, and $(a_i^-,b_i^-,c_i^-)$ for $i=1,\cdots,I^-$ if $a_i\le 0$ or $b_i\le 0$. We have $I=I^++I^-$ and $I^+,I^->0$. Then $\mathcal{R}=\bigcap_{i=1}^{I^+}\{(P,D)|a_i^+ P+b_i^+ D \ge c_i^+\}\cap\bigcap_{i=1}^{I^-}\{(P,D)|a_i^- P+b_i^- D \ge c_i^-\}$. For $1\le l \le I^+$, define $\mathcal{L}_l=\{(P,D)|a_l^+ P+b_l^+ D = c_l^+\}\cap\bigcap_{i=1,i\neq l}^{I^+}\{(P,D)|a_i^+ P+b_i^+ D \ge c_i^+\}\cap\bigcap_{i=1}^{I^-}\{(P,D)|a_i^- P+b_i^- D \ge c_i^-\}$.

For all $(P,D)\in \mathcal{L}_l$, immediately we have $(P,D)\in \mathcal{R}$. For all $(P',D')\in \mathcal{R}$, since $a_l^+ P'+b_l^+ D' \ge c_l^+=a_l^+ P+b_l^+ D$, it should hold that $P'\ge P$ or $D'\ge D$. According to the definition of $\mathcal{L}$, we have $(P,D)\in\mathcal{L}$. Therefore $\mathcal{L}_l \subseteq \mathcal{L}$.

For all $(P,D)\in \mathcal{L}$, we investigate three cases: 1) If $a_i^+ P+b_i^+ D > c_i^+$ for all $1 \le i \le I^+$ and $a_i^- P+b_i^- D > c_i^-$ for all $b_i^->0$, set $\epsilon=\min_{b_i>0}\frac{a_i P+b_i D-c_i}{b_i}$ so that $a_i P+b_i (D-\epsilon) \ge c_i$ for all $b_i>0$. Since $(P,D)\in\mathcal{R}$, for all $b_i\le 0$ $a_{i} P+b_{i} D \ge c_i$, therefore $a_{i} P+b_{i} (D-\epsilon) \ge c_i$ for all $b_i\le 0$. Hence $(P,D-\epsilon)\in\mathcal{R}$, which is against the definition of $\mathcal{L}$. 2) If $a_i^+ P+b_i^+ D > c_i^+$ for all $1 \le i \le I^+$ and $a_i^- P+b_i^- D > c_i^-$ for all $a_i^->0$, set $\epsilon=\min_{a_i>0}\frac{a_i P+b_i D-c_i}{a_i}$ so that $a_i (P-\epsilon)+b_i D \ge c_i$ for all $a_i>0$. Since $(P,D)\in\mathcal{R}$, for all $a_i\le 0$ $a_{i} P+b_{i} D \ge c_i$, therefore $a_{i} (P-\epsilon)+b_{i} D \ge c_i$ for all $a_i\le 0$. Hence $(P-\epsilon,D)\in\mathcal{R}$, which is against the definition of $\mathcal{L}$. 3) If $a_i^+ P+b_i^+ D > c_i^+$ for all $1 \le i \le I^+$, and there exists $i^*$ and $j^*$ such that $a_{i^*}^-\le0$, $b_{i^*}^->0$, $a_{j^*}^->0$, $b_{j^*}^-\le0$, $a_{i^*}^- P+b_{i^*}^- D = c_{i^*}^-$, $a_{j^*}^- P+b_{j^*}^- D = c_{j^*}^-$. For all $(P',D')\in \mathcal{R}$, either $P'\ge P$, $D'\ge D$ or $P'\le P$, $D'\le D$. If there exists $P'<P$ and $D'<D$, then $(P,D)$ is against the definition of $\mathcal{L}$. If $P'\le P$ and $D'\le D$ for all $(P',D')$, since for all $1 \le i \le I^+$, we have $a_i^+ P+b_i^+ D > c_i^+$, therefore $a_i^+ P'+b_i^+ D' > c_i^+$. Hence $\mathcal{L}_i\cap\mathcal{R}=\emptyset$, which is against the condition. From the above three cases, for all $(P,D)\in \mathcal{L}$, there exists at least one certain $l^*$ such that $a_{l^*}^+ P+b_{l^*}^+ D = c_{l^*}^+$, which means $(P,D)\in\mathcal{L}_{l^*}$.

From above we can see that $\mathcal{L}=\bigcup_{l=1}^{I^+}\mathcal{L}_l$. Therefore $\mathcal{L}$ is piecewise linear.

\textbf{Properties of Vertices of $\mathcal{L}$:}

The vertices of $\mathcal{L}$ are also the vertices of $\mathcal{R}$, and neighbouring vertices of $\mathcal{L}$ are also neighbouring vertices of $\mathcal{R}$. From the results in Theorem \ref{theorem_region=polygon}, vertices of $\mathcal{L}$ are obtained by deterministic scheduling policies, and the policies corresponding to neighbouring vertices of $\mathcal{L}$ are different in only one state.

\label{proof_piecewiselinear}

\footnotesize
\bibliographystyle{IEEEtran}
\bibliography{IEEEabrv,scheduling}
\normalsize

\end{document}